\documentclass[11pt]{article}

\usepackage{opex3}
\usepackage{graphics}
\usepackage{graphicx}
\usepackage{epsfig}
\usepackage{amsthm,amsmath,amssymb,amsfonts,bm,subfigure}
\usepackage{fancyhdr}

\usepackage{graphicx}

\definecolor{MyGray}{rgb}{0.92,0.93,0.94}

\newcommand{\dft}[1] {\mathcal{F}\{#1\}}
\newcommand{\dftx}[1] {\mathcal{F}_{\bx}\{#1\}}

\newcommand{\hF} {\hat{F}}
\newcommand{\hpsi} {\hat{\psi}}

\newcommand{\bx} {{\bf x}}

\newcommand{\br} {{\bf r}}
\newcommand{\Diag}[1] {\mbox{Diag}\left( #1 \right)}
\newcommand{\barf}{\bar{f}}
\newcommand{\barb}{\bar{b}}
\newcommand{\partf}{\partial f}
\newcommand{\partpsi}{\partial \psi}
\newcommand{\partbpsi}{\partial \bar{\psi}}

\newcommand{\barpsi}{\bar{\psi}}
\newcommand{\barphi}{\bar{\phi}}
\newcommand{\barz}{\bar{z}}
\newcommand{\yell}{y^{(\ell)}}
\newcommand{\yellu}{y^{(\ell+1)}}
\newcommand{\yelld}{y^{(\ell+2)}}

\begin{document}
\fancyhead[]{}
\fancyfoot[RO]{LBNL-4598E}

\pagestyle{fancy}

\title{Iterative Algorithms for Ptychographic Phase Retrieval}
\author{Chao Yang}
\address{Computational Research Division,
Lawrence Berkeley National Laboratory, Berkeley, CA 94720.}
\email{cyang@lbl.gov}
\author{Jianliang Qian}
 \address{Department of Mathematics, Michigan State University, 
East Lansing, MI.} 
\email{qian@math.msu.edu}
\author{Andre Schirotzek}
\address{Advanced Light Source, Lawrence Berkeley National 
  Laboratory, Berkeley, CA 94720.}
\email{aschirotzek@lbl.gov}
\author{Filipe Maia}
\address{NERSC, Lawrence Berkeley National Laboratory, Berkeley, CA 94720.}
\email{frmaia@lbl.gov}
\author{Stefano Marchesini}
\address{Advanced Light Source, Lawrence Berkeley National Laboratory, 
Berkeley, CA 94720.}
\email{smarchesini@lbl.gov}

\date{\today}
\begin{abstract}
  Ptychography promises diffraction limited resolution
  without the need for high resolution lenses. To achieve high
  resolution one has to solve the phase problem for many partially
  overlapping frames. Here we review some of the existing methods for
  solving ptychographic phase retrieval problem from a numerical
  analysis point of view, and propose alternative methods based on
  numerical optimization.
\end{abstract}

\bibliographystyle{unsrt}
\bibliography{ptyco}

\section{Introduction}
An emerging imaging technique in X-ray science is to use
a localized moving probe to collect multiple diffraction measurements
of an unknown object
\cite{Hegerl1970,Nellist1995, Chapman1996, pie1,pie2, Rodenburg2008, fienup, thibault}. This technique is called ``ptychography".
In a ptychography experiment, one collects a sequence of diffraction
images of dimension $m \times m$. Each image frame $y_{\bx}(\br')$ represents
the magnitude of the Fourier transform of $a(\br)\hpsi(\br+\bx)$,
where $a(\br)$ is a localized illumination (window) function or a probe, 
$\hpsi(\br)$ is the unknown object of interest, and $\bx$ is a 
translational vector. We can express $y_{\bx}$ as 
\begin{equation}
y_{\bx}(\br') = | \dft{a(\br)\hpsi(\br+\bx)} |, \label{data}
\end{equation}
where $\dft{f}$ denotes the Fourier transform of $f$ with respect to $\br$.

In order to reconstruct the unknown object, we must retrieve the 
phases of the measured images.  A number of methods have been proposed to 
recover $\hpsi(\br)$ from ptychographic measurements $y_{\bx}(\br')$ 
\cite{bates-rodenburg,pie1,pie2,fienup,thibault}.
The connection among these methods is not entirely clear from the existing 
literature. Furthermore, little detail is provided on the 
convergence rate or computational efficiency of these methods.

In this paper, we review some of the existing methods
for solving ptychographic phase retrieval problem from a numerical
analysis point of view, and propose to solve the problem 
by alternative methods that are standard in the numerical
optimization community.  In particular, we formulate the ptychographic
phase retrieval problem as an unconstrained nonlinear minimization
problem in section~\ref{sec:opt}, and compare the convergence of several 
well known iterative methods for solving this type of problem in 
section~\ref{sec:example}.  
We discuss computational details such as line search and preconditioning 
that are important for achieving optimal performance in these methods 
in section~\ref{sec:opt}.  We also describe the connection between 
optimization based algorithms and projection algorithms that are often 
discussed in the phase retrieval literature in section~\ref{fixptalg}.

We point out that ptychographic minimization problem is 
not globally convex, which means that iterative methods can be 
trapped at a local minimizer if a poor starting guess is chosen.
We show by a numerical example that one way to escape from
a local minimizer is to switch to a different objective function
in section~\ref{sec:example}.

We observed that the convergence of the optimization based iterative 
algorithms used to perform ptychographic phase retrieval is accelerated 
when the amount of overlap between two adjacent image frames increases. 
We provide an intuitive explaination on why the amount of overlap
between adjacent frames affects the convergence of iterative 
optimization algorithms in section~\ref{sec:example}.

An alternative approach for performing ptychographic phase retrieval
is a method known as Wigner deconvolution.  We review this approach
in section~\ref{sec:wigner} and point out its connection to iterative 
algorithms and its limitations.

We use standard linear algebra notation whenever possible to
describe various quantities evaluated in the iterative algorithms
we present.  To simplify notation we use $a/b$ to denote an element-wise 
division between two vectors $a$ and $b$. Similarly, we use
$a \cdot b$ to denote an element-wise multiplication of $a$ and $b$.
We also use $a^2$ and $a^{1/2}$ occasionally to denote 
the element-wise square and square root of $a$ respectively.
The conjugate of a complex variable $a$ is denoted by $\bar{a}$.  
The real part of $a$ is denoted by $\mbox{Re}(a)$. The conjugate 
transpose of a matrix (or a vector) $A$ is denoted by $A^{\ast}$.
The $|x|$ symbol is reserved for the magnitude (or absolute value) of $x$. 
The Euclidean norm of $x$ is denoted by $\| x \| = \sqrt{x^{\ast} x}$.
We use $\Diag{x}$ to represent a diagonal matrix with the vector $x$ 
on its diagonal.

\section{Ptychographic reconstruction formulated as an optimization problem}
\label{sec:opt}
The problem we would like to solve is to recover $\hpsi$ from a number of
intensity measurements represented by (\ref{data}).  For a finite set
of translational vectors $\bx_i$, we will denote each measurement
by 
\[
b_i = |F Q_i \hpsi|, \ \ i = 1, 2,..., k,
\]
where $\hpsi$ is the sampled unknown object that contains $n$ pixels,
$b_i$ is a sampled measurement that contains $m$ pixels, $F$ is the
matrix representation of a discrete Fourier transform, and $Q_i$ is an
$m \times n$ ``illumination matrix'' that extracts a frame containing
$m$ pixels out of an image containing $n$ pixels.  Each row of $Q_i$
contains at most one nonzero element.  The nonzero values in $Q_i$ are
determined by the illumination function $a(\br)$.

\begin{figure}[hbtp]
  \begin{center}
    \includegraphics[width=0.9\textwidth]{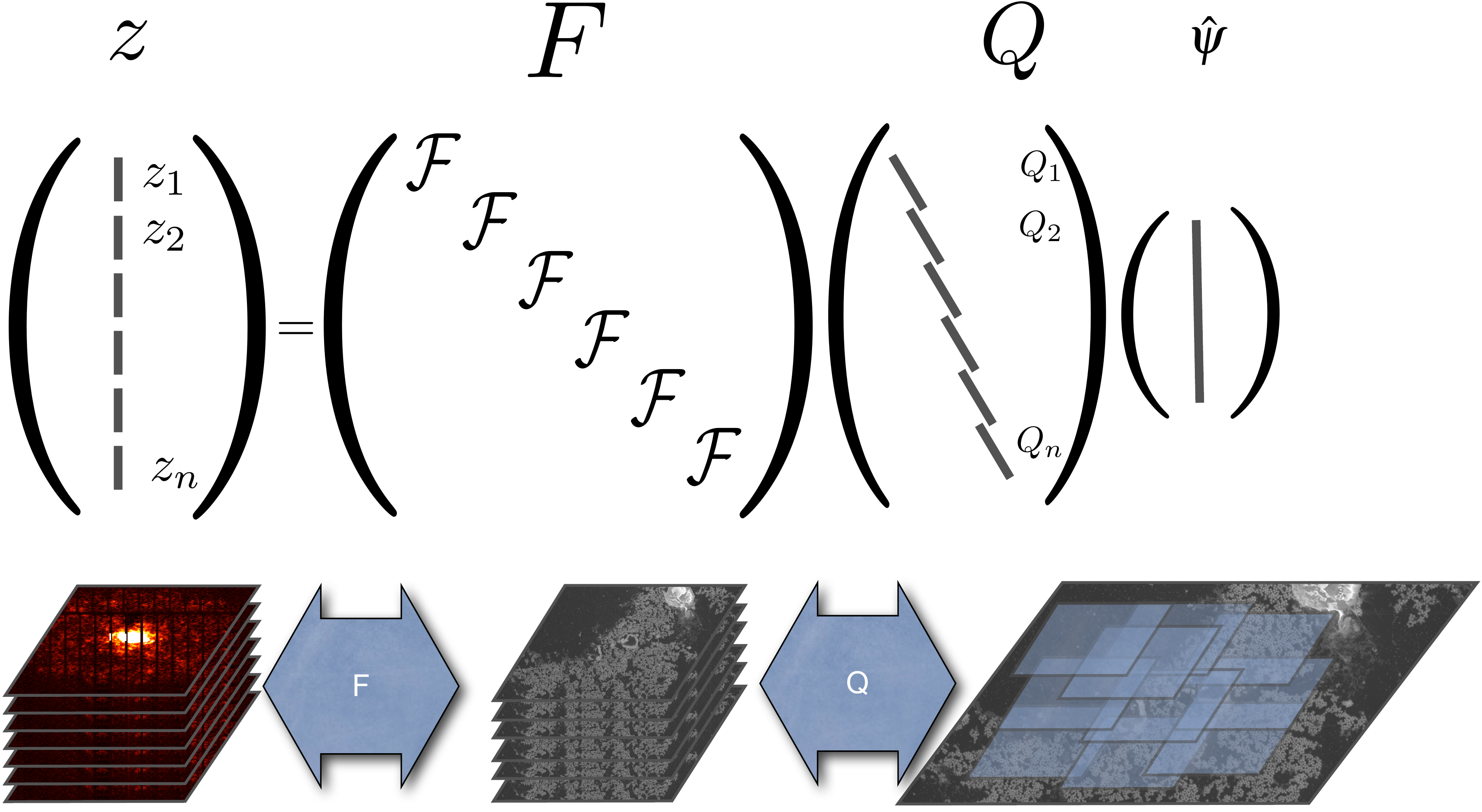}
  \end{center}
  \caption{An unknown object of interest $\hpsi$, and the measured amplitudes $z_x$ related by matrix operations}
  \label{fig:eq1}
\end{figure}

Given a set of measurements, $b_1$, $b_2$, ..., $b_k$, we may attempt
to recover $\hpsi$ by solving the least squares problem
\begin{equation}
  \min_{\psi} \frac{1}{2}\sum_{i=1}^k \| |z_i| - b_i \|^2,
  \label{obj1}
\end{equation}
where $z_i \equiv FQ_i\psi$, and the factor of $1/2$ is included
here merely for convenience.

An alternative objective function we may use to recover $\hat{\psi}$
is
\begin{equation}
\epsilon =  \frac{1}{2}\sum_{i=1}^k \| |z_i|^2 - b_i^2 \|^2,
\label{obj2}
\end{equation}
where $|z_i|^2$ and $b_i^2$ denote vectors obtained from squaring each 
component of $|z_i|$ and $b_i$ respectively.  The advantage of using
(\ref{obj2}) is that it is more differentiable, hence more amenable to
analysis.  In practice, we found the objective function in 
(\ref{obj1}) to be a better choice in terms of computational efficiency
in most cases.

To obtain the minimizers of (\ref{obj1}) or (\ref{obj2}) using 
numerical optimization techniques, we often need to evaluate the
gradient and possibly the Hessian of these objective functions.
Because both (\ref{obj1}) and (\ref{obj2}) are real-valued functions 
of a (potentially) complex vector $\psi$, derivative calculations must
be done with care. One can either take the derivative of (\ref{obj1})
and (\ref{obj2}) with respect to the real and imaginary parts of 
$\psi$ independently or follow the $\mathbb{CR}$-calculus formalism
established in \cite{remmert,delgado} by treating $\psi$ and $\barpsi$ 
as two independent variables.  The latter approach is what we use throughout this
paper.

\subsection{Gradient}
If we let $r_i \equiv |z_i|^2 - b_i^2$, and define
\[
r \equiv \left(
\begin{array}{c}
r_1 \\
r_2 \\
\vdots \\
r_k
\end{array}
\right),
\]
we can rewrite (\ref{obj2}) as $\epsilon(\psi) = r^T r/2$.
Let the matrix $J_i = \partial r_i / \partial \psi$ be the Jacobian of $r_i$ 
with respect to $\psi$.  It follows from the chain rule that
the gradient of $\epsilon$ in vector form is
\begin{equation}
\nabla \epsilon(\psi) = \biggl(\frac{\partial{\epsilon}}{\partial \psi} \biggr)^{\ast}
= \biggl(\frac{\partial \epsilon}{\partial r} 
\frac{\partial r}{\partial \psi} \biggr)^{\ast}
= J^{\ast} r, 
\label{jr}
\end{equation}
where
\[
J \equiv \left(
\begin{array}{c}
J_1 \\
J_2 \\
\vdots \\
J_k
\end{array}
\right).
\]

Note that we may rewrite $|z_i|^2$ as
$\mbox{Diag}(z_i)^{\ast}z_i$, where $\mbox{Diag}(x)$ denotes
a diagonal matrix that contains the vector $x$ on its diagonal
and $z_i \equiv FQ_i \psi$.  Using this observation, we can show that
\begin{equation}
J_i = \frac{\partial |z_i|^2}{\partial \psi} 
=\frac{\partial \Diag{\bar{z}_i} z_i}{\partial z_i} \frac{\partial z_i}{\partial \psi}
+ \frac{\partial \Diag{z_i} \bar{z}_i}{\partial \bar{z}_i}
\frac{\partial \bar{z}_i}{\partial \psi}
= \mbox{Diag}(\bar{z}_i)FQ_i = \mbox{Diag}(FQ_i\psi)^{\ast}FQ_i.
\label{jac1}
\end{equation}
It follows from (\ref{jr}) and the above expression that
\begin{equation}
\nabla \epsilon = \sum_{i=1}^k Q_i^{\ast} F^{\ast} 
\mbox{Diag}(z_i)[|z_i|^2 - b_i^2]. \label{egrad}
\end{equation}

The gradient of the objective function in (\ref{obj1}), which we will
denote by $\rho(\psi)$, is slightly more complicated.  By rewriting
$|z_i|$ as $(|z_i|^2)^{1/2}$, with the understanding that the square
root is taken component-wise, and by using the chain rule and replacing
${\partial |z_i|^2}/{\partial \psi}$ with the expression given in
(\ref{jac1}), we obtain
\[
J_i = \frac{\partial |z_i|}{\partial \psi} =
\frac{\partial (|z_i|^2)^{1/2}}{\partial |z_i|^2} \cdot 
\frac{\partial |z_i|^2}{\partial \psi} 
= \frac{1}{2} \Diag{\barz_i/|z_i|} F Q_i,
\]
if $|z_i|$ does not contain any zero element for all $i = 1,2,...,m$.

Consequently, we may now express $\nabla \rho(\psi)$ as
\begin{eqnarray}
\nabla \rho(\psi) = J^{\ast}r  &=& \sum_{i=1}^k J_i^{\ast}r_i \nonumber \\
&=& \frac 1 2 \sum_{i=1}^k   Q_i^\ast F^\ast \Diag{\frac{z_i}{|z_i|}}
(|z_i|-b_i) \nonumber \\
&=&
 \frac{1}{2}
\sum_{i=1}^k Q_i^{\ast}F^{\ast}\left [z_i-\Diag{\frac{z_i}{|z_i|}}b_i 
\right ]
\nonumber \\
&=&\frac{1}{2}\sum_{i=1}^k Q_i^{\ast}F^{\ast}\left[FQ_i\psi
-\Diag{\frac{z_i}{|z_i|}} b_i\right] \nonumber \\
&=& \frac{1}{2}\sum_{i=1}^k \left[Q_i^{\ast}Q_i\psi -Q_i^{\ast}F^{\ast}
\Diag{\frac{z_i}{|z_i|}} b_i \right]\label{gradshort}.
\end{eqnarray}
Recall that $z_i = FQ_i\psi$. Thus, the 
expression $\mbox{Diag}(FQ_i\psi)\mbox{Diag}(|z_i|)^{-1} b_i$ simply
represents projecting $\psi$ onto the Fourier magnitude constraint imposed
by the data $b_i$.  Note that the expression given above for the 
gradient of $\rho(\psi)$ is only valid when $|z_i|$ does not contain any zero
for all $i=1,2,...,m$. If $|z_i|$ contains a zero component for some $i$, 
and if the corresponding component in $b_i$ is nonzero,
$\nabla \rho$ is not well defined, i.e., $\nabla \psi$ has singularities
at $\psi$'s where $FQ_i\psi$ contains a zero element for some $i$.

Note that both (\ref{egrad}) and (\ref{gradshort}) 
remain real when $\psi$ is real and when $b_i$ is obtained from 
a discrete Fourier transform of a real image (so that conjugate 
symmetry is preserved in $\Diag{z_i/|z_i|} b_i$.)

The directional derivatives of $\epsilon$ and $\rho$ along a 
direction $\phi$ are defined by
\begin{equation}
\frac{\partial \epsilon}{\partial \psi} \phi + \frac{\partial \epsilon}{\partial \barpsi} \bar{\phi} 
= 2 \sum_{i=1}^k \mbox{Re} \biggl[(|z_i|^2 - b_i^2)^T \Diag{z_i}^{\ast}FQ_i \phi \biggr]
\label{dirdiff1}
\end{equation}
and
\begin{equation}
\frac{\partial \rho}{\partial \psi} \phi + \frac{\partial \rho}{\partial \barpsi} \bar{\phi} \\
= \sum_{i=1}^k \mbox{Re} \biggl[\phi^{\ast}Q_i^{\ast}Q_i\psi-\phi^{\ast}
Q_i^{\ast}F^{\ast}\Diag{\frac{z_i}{|z_i|}} b_i \biggr]
\label{dirdiff2}
\end{equation}
respectively

\subsection{Hessian} \label{sec:hess}
The Hessian of $\epsilon(\psi)$ and $\rho(\psi)$ provides information
on the convexity of these objective functions.  A globally convex 
function has a unique minimizer. Such a minimizer can be obtained by
standard optimization techniques that we will describe in the next 
section.  If the objective function is not convex everywhere, a standard
optimization algorithm may produce a local minimizer that is not the
true solution to ptychographic reconstruction problem.

Again, because both $\epsilon(\psi)$ and $\rho(\psi)$ are real valued
functions of a potentially complex vector $\psi$, their Hessians are
defined as
\[
H^{o} = \left(
\begin{array}{cc}
H_{\psi\psi}^o    & H_{\psi\barpsi}^o \\
H_{\barpsi\psi}^o & H_{\barpsi\barpsi}^o
\end{array}
\right),
\]
where 
\[
H_{\psi\psi}^o \equiv \frac{\partial}{\partpsi}\biggl(\frac{\partf}{\partpsi}\biggr)^{\ast}, \ \
H_{\barpsi\psi}^o \equiv \frac{\partial}{\partbpsi}\biggl(\frac{\partf}{\partpsi}\biggr)^{\ast}, \ \
H_{\psi\bar\psi}^o \equiv \frac{\partial}{\partpsi}\biggl(\frac{\partf}{\partbpsi}\biggr)^{\ast}, \ \
H_{\barpsi\barpsi}^o \equiv \frac{\partial}{\partbpsi}\biggl(\frac{\partf}{\partbpsi}\biggr)^{\ast},
\]
and $o$ is either $\epsilon$ or $\rho$.

It is not difficult to show  that
\begin{eqnarray}
H_{\psi\psi}^{\epsilon} &=& \sum_i Q_i^{\ast}F^{\ast} \Diag{2|z_i|^2 - b_i^2} FQ_i \label{hpsihpsi1}, \\
H_{\barpsi\barpsi}^{\epsilon}  &=& \sum_i Q_i^T F^T \Diag{2|z_i|^2 - b_i^2} \bar{F}\bar{Q}_i \label{hbpsibpsi1}, \\
H_{\psi\barpsi}^{\epsilon}  &=& \sum_i Q_i^{\ast}F^{\ast}  \Diag{z_i}^2 \bar{F} \bar{Q}_i \label{hpsibpsi1}, \\
H_{\barpsi\psi}^{\epsilon}  &=& \left(H_{\psi\barpsi}^{\epsilon}\right)^{\ast} = \sum_i Q_i^T F^T \Diag{\bar{z}_i}^2 F Q_i. \label{hbpsipsi1}
\end{eqnarray} 

If we let $t_{ji} \equiv |t_{ji}|e^{i\mu_{ji}}$, $\zeta_{ji} \equiv |\zeta_{ji}|e^{i\theta_{ji}}$ 
and $\beta_{ji}$ be the $j$th component of 
$t_i = FQ_i\phi$, $z_i=FQ_i\psi$ and $b_i$ respectively, then the 
curvature $\tau_\epsilon(\psi,\phi)$ at $\psi$ along any direction $\phi$ 
can be calculated as follows
\begin{eqnarray}
\tau_\epsilon(\psi,\phi) &=& (\phi^{\ast} \ \ \phi^T) 
\left(
\begin{array}{cc}
H_{\psi\psi}^\epsilon     & H_{\psi \barpsi}^\epsilon \\
H_{\barpsi \psi}^\epsilon & H_{\barpsi \barpsi}^\epsilon
\end{array}
\right)
\left(
\begin{array}{c}
\phi \\
\bar{\phi}
\end{array}
\right) \nonumber \\
&=&\sum_i (t_i^{\ast} \ \ t_i^T) 
\left(
\begin{array}{cc}
 \Diag{2|z_i|^2 - b_i^2}    &  \Diag{z_i}^2\\
\Diag{\bar z_i}^2 &  \Diag{2|z_i|^2 - b_i^2} 
\end{array}
\right)
\left(
\begin{array}{c}
t_i \\
\bar{t_i}
\end{array}
\right) \nonumber  \\
&=& \sum_i
2 t_i^{\ast} \Diag{2|z_i|^2 - b_i^2} t_i 
+ 2 \mbox{Re}[t_i^T \Diag{\bar{z}_i}^2 t_i] \nonumber \\
&=& \sum_i
2 t_i^{\ast} \Diag{|z_i|^2 - b_i^2} t_i + 2\biggl( t_i^{\ast}\Diag{|z_i|}^2 t_i 
+ \mbox{Re}[t_i^T \Diag{\bar{z}_i}^2 t_i] \biggr) \nonumber \\
&=& 2 \sum_{i=1}^k \sum_{j=1}^n
 |t_{ji}|^2(|z_{ji}|^2 - \beta_{ji}^2) + \biggl(|t_{ji}|^2 |z_{ji}|^2
+ \mbox{Re}\biggl[(t_{ji}\barz_{ji})^2\biggr]\biggr) \nonumber \\
&=& 2 \sum_{i=1}^k \sum_{j=1}^n
 |t_{ji}|^2(|z_{ji}|^2 - \beta_{ji}^2) + 2|t_{ji}|^2 |z_{ji}|^2 \cos^2 (\mu_{ji} - \theta_{ji}). 
\label{curv1}
\end{eqnarray}
At the minimizer of $\epsilon(\psi)$, $|z_i| = b_i$. So the first term
of (\ref{curv1}) is zero.  Because the second term of 
(\ref{curv1}) is nonnegative, $\tau \geq 0$, i.e., $\epsilon$ is convex at 
the solution.  Moreover, the convexity of $\epsilon$ is preserved 
in the area where $|z_{ji}| \geq \beta_{ji}$.

A similar observation can be made from the curvature of $\rho$. It is 
not difficult to show  that
\begin{eqnarray}
H_{\psi\psi}^\rho &=& \frac{1}{2}\left(\sum_{i=1}^k Q_i^{\ast}Q_i - \frac{1}{2}
Q_i^{\ast}F^{\ast}\Diag{\frac{b_i}{|z_i|}} FQ_i\right), \label{hpsipsi2} \\
H_{\barpsi \barpsi}^\rho &=&\frac{1}{2}\left(\sum_{i=1}^k Q_i^T \bar{Q}_i
-\frac{1}{2} Q_i^TF^T \Diag{\frac{b_i}{|z_i|}} \bar{F}\bar{Q}_i\right), \label{hbpsibpsi2} \\
H_{\psi\barpsi}^\rho &=& \frac{1}{4} \sum_{i=1}^k Q_i^{\ast}F^{\ast} 
\Diag{\frac{b_i \cdot z_i^2}{|z_i|^3}} \bar{F}\bar{Q}_i, \label{hpsibpsi2} \\
H_{\barpsi\psi}^\rho &=& \frac{1}{4} \sum_{i=1}^k Q_i^{T}F^{T} 
\Diag{\frac{b_i \cdot \barz_i}{|z_i|}} FQ_i. \label{hbpsipsi2}
\end{eqnarray}
It follows that
\begin{eqnarray}
\tau_{\rho}(\psi,\phi) &=& (\phi^{\ast} \ \ \phi^T)
\left(
\begin{array}{cc}
H_{\psi\psi}^\rho    & H_{\psi\barpsi}^\rho \\
H_{\barpsi\psi}^\rho & H_{\barpsi\barpsi}^\rho
\end{array}
\right)
\left(
\begin{array}{c}
\phi \\
\barphi
\end{array}
\right)\\
&=&\nonumber  \frac 1 2\sum_i
 (t_i^{\ast} \ \bar t_i^T)
\left(
\begin{array}{cc}
I - \frac{1}{2}\Diag{\tfrac{b_i}{|z_i|}}
& \frac{1}{2}\Diag{\tfrac{b_i}{|z_i|} \cdot \tfrac{z_i^2}{|z_i|^2}}
\\
\frac{1}{2}\Diag{\tfrac{b_i}{|z_i|} \cdot \tfrac{\bar z_i^2}{|z_i|^2}} &
I-\frac{1}{2}\Diag{\tfrac{b_i}{|z_i|}}
\end{array}
\right)
\left(
\begin{array}{c}
t_i \\
\bar{t}_i
\end{array}
\right)
\nonumber \\
&=&  \frac{1}{2}\sum_{i=1}^k\sum_{j=1}^n \biggl(
 2 |t_{ji}|^2 - |t_{ji}|^2 \frac{\beta_{ji}}{|\zeta_{ji}|} +  \mbox{Re} 
\biggl[\bar{t}_{ji}^2 
\frac{\beta_{ji} \zeta_{ji}^2}{|\zeta_{ji}|^3}\biggr]\biggr) \nonumber \\
&=&  \frac{1}{2}\sum_{i=1}^k\sum_{j=1}^n |t_{ji}|^2\biggl(
 2 - \frac{\beta_{ji}}{|\zeta_{ji}|}+  \frac{\beta_{ji}}{|\zeta_{ji}|}\mbox{Re} \biggl[
\frac{\bar{t}_{ji}^2}{|t_{ji}|^2}\frac{\zeta_{ji}^2}{|\zeta_{ji}|^2}\biggr]\biggr) \nonumber \\
&=&  \sum_{i=1}^k\sum_{j=1}^n |t_{ji}|^2\left(
 1 - \frac{\beta_{ji}}{|\zeta_{ji}|} \sin^2(\mu_{ji}-\theta_{ji}) \right).
\label{curv2}
\end{eqnarray}
Thus, $\tau_\rho \geq 0$ when $|\zeta_{ji}| \geq \beta_{ji}$ for all
$j=1,2,...,n$ and $i = 1, 2, ...,k$. Even if $|\zeta_{ji}|$ is
slightly less than $\beta_{ji}$ for some $j$ and $i$, $\tau_\rho$ may
remain positive when the corresponding $\sin^2(\mu_{ji}-\theta_{ji})$
is sufficiently small and other terms in the summation in
(\ref{curv2}) are sufficiently large and positive.  

A typical problem
encountered in optics is when $k=1$. When only one diffraction image
is recorded, experience shows that local minima are common.  
Regions of negative curvature separate local minima from the global solution. 

\section{Iterative Algorithms based on Nonlinear Optimization} \label{sec:algs} Because the gradient
and Hessian of (\ref{obj1}) and (\ref{obj2}) are relatively easy to
evaluate, we may use standard minimization algorithms such as the
steepest descent method, the Newton's method and the nonlinear
conjugate gradient method to find the solution to the ptychographic
reconstruction problem.  We will review some of these methods in
section~\ref{optalgs} and discuss some techniques for improving
the performance of these algorithms in the rest of this section.

\subsection{Basic Algorithms} \label{optalgs}
In many standard numerical optimization algorithms, we construct a sequence of 
approximations to $\hpsi$ by
\begin{equation}
\psi^{(\ell+1)} = \psi^{(\ell)} + \beta p^{(\ell)},
\label{optupdate}
\end{equation}
where $p^{(\ell)}$ is a {\em search direction} along which
the objective function (\ref{obj1}) or (\ref{obj2}) decreases, and 
$\beta > 0$ is an appropriate step length.  

The simplest type of search direction is the {\em steepest descent} direction
\[
p_{\mathrm{sd}}^{(\ell)} = -\nabla_{\psi} o(\psi^{(\ell)},\barpsi^{(\ell)}),
\]
where $o$ is either $\epsilon$ or $\rho$.
When the Hessian of $\rho$ or $\epsilon$ is positive definite at 
$\psi^{(\ell)}$, the Newton's direction $p_{\mathrm{nt}}^{(\ell)}$,
which is the solution of 
\begin{equation}
\left(
\begin{array}{cc}
H_{\psi\psi}^{o}    & H_{\psi\barpsi}^{o} \\
H_{\barpsi\psi}^{o} & H_{\barpsi\barpsi}^{o} \end{array}
\right)
\left(
\begin{array}{c}
p_{\mathrm{nt}}^{(\ell)}  \\
\bar{p}_{\mathrm{nt}}^{(\ell)}  
\end{array}
\right)
=
\left(
\begin{array}{c}
p_{\mathrm{sd}}^{(\ell)}  \\
\bar{p}_{\mathrm{sd}}^{(\ell)}
\end{array}
\right),
\label{newtondir}
\end{equation}
is also a descent direction.

Due to the nonlinear least squares nature of the objective functions
(\ref{obj1}) and (\ref{obj2}), we may replace the true Hessian 
in (\ref{newtondir}) by a simpler approximation constructed from 
the Jacobian of the residual functions $|z_i| - b_i$ or $|z_i|^2 - b_i^2$
for $i = 1,2,...,k$. This technique yields the {\em Gauss-Newton} (GN) 
search directions.

Both Newton's method and the Gauss-Newton method require solving 
a system of linear equations at each step in order to obtain a search 
direction. Because the dimension of these linear systems is $n \times n$,
where $n$ is the number of pixels in the image to be reconstructed,
constructing the Hessian or Jacobian and
solving these equations by matrix factorization based methods 
will be prohibitively expensive.  Iterative methods that make use 
of matrix vector multiplications without forming the Hessian or
the $J$ matrix explicitly are more appropriate.  However, 
several iterations may be required to reach a desired accuracy
needed to produce a good search direction. Hence methods based on 
Newton or Gauss-Newton search directions tend to be more expensive.

The Hessian in (\ref{newtondir}) can also be 
replaced by approximations
constructed from changes in the gradient computed at each iteration. 
Such approximation yields {\em Quasi-Newton} search directions.

Another commonly used search direction is the conjugate gradient direction
defined by
\[
p_{\mathrm{cg}}^{(\ell)} = -g^{(\ell)} + \alpha p_{\mathrm{cg}}^{(\ell-1)},
\]
where $g^{(\ell)}$ is the gradient of (\ref{obj1}) or (\ref{obj2}) at
$\psi^{(\ell)}$ and $\alpha$ is often chosen to be
\[
\alpha = \frac{\mbox{Re}\left[(g^{(\ell)})^{\ast}(g^{(\ell)} - g^{(\ell-1)})\right]}{\|g^{(\ell-1)}\|^2}.
\]
This choice of $\alpha$ yields what is known as the {\em Polak-Ribiere} conjugate gradient method.

There are a variety of ways to choose an appropriate step length $\beta$ in
(\ref{optupdate}). They are often referred to as {\em line search} methods.
The purpose of line search is to ensure that the objective function
decreases as we move from $\psi^{(\ell)}$ to $\psi^{(\ell+1)}$ so that
$\psi^{(\ell)}$ will converge to at least a local minimizer as $\ell$ 
increases. Such type of convergence is often called {\em global convergence}.

Another way to achieve global convergence in an iterative optimization 
procedure is to use the {\em trust region} technique to determine 
a search direction and step length simultaneously. Under the trust
region framework, we minimize the true objective function by
minimizing a sequence of simpler ``surrogate" functions that mimic the
behavior of the true objective function within a small neighborhood
of the current approximations. That is, in each step of this iterative
optimization procedure, we solve what is called a trust region subproblem
\begin{equation}
\min_{\| \phi \| \leq \Delta} q(\psi^{(\ell)}+\phi),
\label{trsub}
\end{equation}
where $q(\psi)$ is the surrogate function, and the parameter
$\Delta$ is known as a trust region radius that defines the
region in which $q(\psi)$ approximates $\rho(\psi)$ or $\epsilon(\rho)$ well.
Such a radius must be chosen carefully.  It may need to be adjusted 
iteratively based on the ratio of the reduction in the surrogate and 
the reduction in the true objective function achieved by the solution
to (\ref{trsub}).

A commonly used surrogate function is the second order Taylor expansion of 
the true objective function.  The minimizer of this particular surrogate 
gives a full step Newton direction.  However, the Newton step may not 
satisfy the trust region constraint, thus may not be the solution to
(\ref{trsub}).

The trust region subproblem can be solved either exactly or approximately
depending on the cost of evaluating $q(\psi)$ and its derivatives. 
If the second order Taylor expansion is chosen as the surrogate, most 
methods need to solve the Newton equation
\[
\nabla^2 q(\phi) s = -\nabla q(\phi),
\]
where $\nabla^2 q$ is the Hessian of the true objective at the 
current iterate $\psi$.  This equation can be solved approximately 
by the (linear) conjugate gradient algorithm when $\nabla^2$ is
positive definite. When $\nabla^2 q$ is not positive definite, 
(\ref{trsub}) can also be solved by following a negative curvature
direction to the boundary of the trust region.  These techniques
are used in an efficient iterative procedure for solving a large-scale
trust region subproblem developed by Steihaug \cite{steihaug}. 
The method requires compute the matrix vector product $\nabla^2 q v$ for
some vector $v$. This product can be approximated by a
finite difference approximation
\[
(\nabla^2 q) v \approx 
\frac{\nabla q(\phi + \eta v) - \nabla(\phi)}{\eta},
\]
for some small $\eta$. Therefore, it is not necessary to explicitly
construct the Hessian of the objective function in Steihaug's method.

\subsection{Weighted Objective and Precondition} \label{sec:prec}
The least squares objective function in (\ref{obj1}) and (\ref{obj2})
can be expressed as
\[
\rho(\psi) = \frac{1}{2} \sum_{i=1}^k\langle |z_i| - b_i, |z_i| - b_i \rangle,
\]
and 
\[
\epsilon(\psi) = \frac{1}{2} \sum_{i=1}^k\langle |z_i|^2 - b_i^2, |z_i|^2 - b_i^2 \rangle
\]
respectively, where $\langle x, y \rangle = x^{\ast} y$ denotes the standard
Euclidean inner product.  This inner product can be replaced by
a weighted inner product $\langle x, y \rangle_B = x^{\ast} B y$, where
$B$ is a symmetric positive definite matrix, to accelerate the 
convergence of iterative methods used to recover the unknown image $\psi$.
As we will show in section~\ref{sec:example}, the choice of
$B = \mbox{Diag}(b_i)^{-1}$ is particularly useful for accelerating
the convergence of all iterative methods we have looked at. To maintain
numerical stability and reduce noise amplification, it is often necessary 
to add a small constant to the diagonal of $B$ to prevent it from becoming
singular or ill-conditioned.

Another useful technique for accelerating iterative methods for 
solving unconstrained minimization problem is preconditioning.  
Instead of minimizing $\rho(\psi)$ or $\epsilon(\psi)$, we make 
a change of variable and minimize $\hat{\rho}(\phi)$ and 
$\hat{\epsilon}(\phi)$, where $\phi = K \psi$, and $K$ is a
preconditioner that is usually required to be Hermitian and positive 
definite.  The purpose of introducing the preconditioner $K$ is to
reduce the condition number of the Hessian of the objective function.
A highly ill-conditioned Hessian often leads to slow convergence of
an iterative method. A well-known example is the zig-zag behavior
of the steepest descent algorithm when it is applied to the 
Rosenbrock function.

It follows from the chain rule and (\ref{gradshort}) that the gradient of 
$\hat{\rho}(\psi)$ is simply
\[
\nabla \hat{\rho}(\psi) = \frac 1 2K^{-1} 
\sum_{i=1}^k [Q_i^{\ast}Q_i\psi -Q_i^{\ast}F^{\ast}\Diag{\frac{z_i}{|z_i|}} b_i ],
\]
where $z_i = FQ_i\psi$.

If we take the preconditioner to be the constant term on the diagonal blocks
of $H_{\psi\psi}^{\rho}$, i.e.,
\begin{equation}
K = \sum_{i = 1}^k Q_i^{\ast}Q_i, \label{precond}
\end{equation}
which is a diagonal matrix, the gradient of $\hat{\rho}$ simply becomes
\[
\nabla \hat{\rho}(\psi) = \frac{1}{2}\left[ \psi - \biggl(\sum_{i=1}^k Q_i^{\ast}Q_i\biggr)^{-1}
\biggl(\sum_{i=1}^k Q_i^{\ast}F^{\ast}\Diag{\frac{z_i}{|z_i|}}b_i\biggr)\right],
\]
and the corresponding preconditioned steepest descent algorithm with 
a constant step length of 2 yields the following updating formula:
\[
\psi^{(\ell+1)} =  \biggl(\sum_{i=1}^k Q_i^{\ast}Q_i\biggr)^{-1}
\biggl(\sum_{i=1}^k Q_i^{\ast}F^{\ast}\Diag{\frac{z_i^{(\ell)}}{|z_i^{(\ell)}|}} b_i\biggr),
\]
where $z_i^{(\ell)} = FQ_i\psi^{(\ell)}$.  This updating formula is
identical to that used in the {\em error reduction} algorithm or 
{\em alternate projection} algorithm mentioned in the standard
phase retrieval literature \cite{Marchesini2007a}, which is guaranteed 
to converge to at least a local minimizer as shown in section~\ref{fixptalg}.

\subsection{Line Search}
The global convergence of an unconstrained optimization algorithm
depends on effective line search strategies.
Assuming that $\phi$ is a descent direction for $\rho(\psi)$ at $\psi$, 
i.e., $\nabla \rho(\psi)^T \phi < 0$, we would like to seek an appropriate 
step length $\alpha$ so that
\[
\rho(\psi + \alpha \phi) < \rho(\psi).
\]

One way to obtain such a step length is to minimize the scalar function
$\xi(\alpha) = \rho(\psi + \alpha \phi)$ with respect to $\alpha$.
This can be done by applying the Newton's method to generate 
a sequence of $\alpha$'s that satisfy
\begin{equation}
\alpha_{i+1} = \alpha_{i+1} - \frac{\xi'(\alpha_i)}{\xi''(\alpha_i)},
\label{newtonalpha}
\end{equation}
and accepting an $\alpha_i$ that satisfies 
\[
\xi(\alpha_i) < c_1 \xi(0), \ \ \mbox{and} \ \ 
|\xi'(\alpha_i)| < c_2 | \xi'(\alpha_{i-1})|,
\]
for some small constants $0 < c_1, c_2 < 1$.  In order to 
obtain the values of $\xi'(\alpha_i)$ and $\xi''(\alpha_i)$ 
required in (\ref{newtonalpha}), we need to evaluate
the directional derivative and curvature of $\rho$ at
$\psi + \alpha_i \phi$ along the search direction $\phi$.
That is,
\begin{eqnarray*}
\xi'(\alpha_i)  &=& 2\mbox{Re}(\phi^{\ast} \nabla \rho(\psi + \alpha_i \phi)) \\
\xi''(\alpha_i) &=& \tau_\rho(\psi + \alpha_i \phi, \phi).
\end{eqnarray*}
Although these derivative calculations will incur additional 
computation, the cost of these computation can be kept at a minimal
by keeping $FQ_i\phi$ in memory as we will discuss at
the end of this section.


We should note that the Newton's method may not always 
succeed in finding an appropriate $\alpha$ due to the fact
that $\xi(\alpha)$ is generally not globally convex.  
The convergence of the Newton's method will depend 
on the choice of the starting guess.  When 
a good starting guess is chosen, we typically need to 
run only a few Newton iterations to reach a reasonable $\alpha$ value. 
Because the purpose of line search is to identify a step length 
that would lead to a sufficient reduction in the objective function,
it is not necessary to find the actual minimizer of $\xi(\alpha)$.

However, an exact line search may not satisfy what is known as the 
second Wolfe condition
\[
\nabla \rho(\psi+\alpha \phi)^{\ast}p \geq c_2 \nabla \rho(\psi)^T \phi,
\]
where $0<c_2<1$ is typically chosen to be 0.9. This condition
on the change of the curvature of the objective function and the
first Wolfe condition
\[
\rho(\psi + \alpha \phi) \leq \rho(\psi) 
+ c_1 \alpha \nabla \rho(\psi)^{\ast} \phi,
\]
for some constant $c_1$ typically chosen to be $10^{-3}$, which is
a condition that guarantees a sufficient decrease in the objective
function, are required to prove the global convergence of
the sequence $\{\psi^{(\ell)}\}$ generated by (\ref{optupdate}) 
in many optimization algorithms.  Line search techniques that satisfy 
both Wolfe conditions
can be found in \cite{morethente} and many other standard optimization
textbooks \cite{nocedalwright}.  We should note that these techniques
may also be sensitive to the choice of the initial guesses to the
step length as well as the choice of $c_1$ and $c_2$ parameters.
When a poor initial guess is chosen, these techniques can yield
$\alpha$ values that are too small.  Strategies for choosing a
good starting guess of $\alpha$ can be found in \cite{nocedalwright} 
also.

Regardless which line search technique one uses, one typically
needs to evaluate the objective function $\epsilon(\psi+\alpha \phi)$ 
or $\rho(\psi+\alpha \phi)$ and its directional derivatives for 
a number of different $\alpha$ values.
If we compute $\tilde{\psi} = \psi+\alpha \phi$ first and 
use the formulae given in (\ref{obj1}), (\ref{obj2}), (\ref{dirdiff1})
and (\ref{dirdiff2}) to evaluate the objective function and 
directional derivative (by replacing $\psi$ with $\tilde{\psi}$),
each evaluation will perform $k$ FFTs.  To reduce the cost of line search,
we may evaluate $t_i = FQ_i \phi$ in advance so that no FFT is required
in the the line search procedure itself. For example, to evaluate (\ref{obj1}),
we can simply compute
\[
\rho(\tilde{\psi}) = \sum_{i=1}^k \left\| |z_i + \alpha t_i| - b_i \right\|^2,
\]
where $z_i = FQ_i\psi$ and $t_i$ have been computed already.
Similarly, the direction derivative of $\rho$ at $\psi + \alpha \phi$ can be
obtained from
\[
\sum_{i=1}^k\mbox{Re}
\left[
t_i^{\ast} (z_i + \alpha t_i) - t_i^{\ast}
\Diag{\frac{z_i + \alpha t_i}{|z_i + \alpha t_i|}} b_i
\right].
\]
Also, notice that no FFT is required in the curvature calculation (\ref{curv2})
once $t_i$'s are available.

\section{Fixed-Point Iteration and Projection Algorithms} \label{fixptalg}
An alternative approach to finding a minimizer of (\ref{obj1}) is to 
set its gradient to zero and seek $\psi$ that satisfies the first
order necessary condition of the minimization problem. If 
$\sum_{i=1}^k Q_i^{\ast}Q_i$ is nonsingular, by setting 
$\nabla \rho(\psi) = \frac{1}{2}\sum_{i=1}^k \left[Q_i^{\ast}Q_i\psi -Q_i^{\ast}F^{\ast} \Diag{\frac{z_i}{|z_i|}} b_i \right]$ to 0, 
we obtain
\begin{equation}
\psi = f(\psi)
\label{fixpoint}
\end{equation}
where 
\begin{equation}
f(\psi) = \biggl(\sum_{i=1}^k Q_i^{\ast}Q_i\biggr)^{-1}
\left[\sum_{i=1}^k Q_i^{\ast}F^{\ast}\Diag{\frac{z_i}{|z_i|}} b_i\right].
\label{fxpoint}
\end{equation}
Recall that $z_i \equiv FQ_i\psi$. Clearly, $\psi$ is a fixed point of the function 
$f$.  

A simple iterative technique one may use to find the solution to (\ref{fxpoint})
is the fixed point iteration that has the form
\[
\psi^{(\ell+1)} = f(\psi^{(\ell)}).
\]
Replacing $f$ with the right hand size of (\ref{fxpoint}) yields
\begin{equation}
\psi^{(\ell+1)} = \left(\sum_{i=1}^k Q_i^{\ast}Q_i\right)^{-1}
\left[\sum_{i=1}^k Q_i^{\ast}F^{\ast}
\Diag{\frac{z_i^{(\ell)}}{|z_i|^{(\ell)}}} b_i\right],
\label{fpupdate}
\end{equation}
where $z_i^{(\ell)} \equiv FQ_i\psi^{(\ell)}$. 
This is the same sequence of iterates produced in what is known
as the {\em error reduction} algorithm in standard phase
retrieval literature \cite{Marchesini2007a}.  This method is also known as the
{\em alternate projection} algorithm for reasons to be 
discussed below.

It is easy to verify that the updating formula in (\ref{fpupdate}) is
identical to that produced by a preconditioned steepest descent
algorithm in which the preconditioner $K$ is chosen to be
$K = \sum_{i=1}^{k} Q_i^{\ast}Q_i$, and a constant step length 
of 2 is taken at each iteration, i.e.,
\[
\psi^{(\ell+1)} = \psi^{(\ell)} - 2 \nabla \rho(\psi^{(\ell)}).
\]

The sequence of iterates $\{\psi^{(\ell)}\}$ produced by (\ref{fpupdate})
is guaranteed to converge to the fixed point of $f(\psi)$ from any 
starting point $\{\psi^{(0)}\}$, if the spectral radius (i.e., the largest
eigenvalue) of the Jacobian of $f$ (with respect to $\psi$) is strictly
less than 1.  Because the function $f$ in (\ref{fixpoint}) can 
be viewed as a function of $\psi$ and $\barpsi$, we should 
examine the Jacobian matrix of the system
\begin{eqnarray}
\psi &=& \left(\sum_{i=1}^k Q_i^{\ast}Q_i\right)^{-1} 
\left[
\sum_{i=1}^k Q_i^{\ast}F^{\ast}\Diag{\frac{z_i}{|z_i|}}b_i
\right],
\label{fixp1} \\
\barpsi &=& (\sum_{i=1}^k Q_i^T\bar{Q}_i)^{-1} 
\left[
\sum_{i=1}^k Q_i^TF^T\Diag{\frac{\bar{z}_i}{|z_i|}}b_i,
\right]
\label{fixp2}
\end{eqnarray}
where (\ref{fixp2}) is simply the conjugate of (\ref{fixp1}). It is 
not difficult to show that this Jacobian matrix has
the form
\begin{equation}
J = 
\left(
\begin{array}{cc}
  K^{-1} & 0 \\
0 &  \bar{K}^{-1} 
\end{array}
\right)
\left(
\begin{array}{cc}
K-2H_{\psi \psi}^{\rho} & -2 H_{\psi \bar \psi}^{\rho} \\
-2H_{\bar \psi \psi}^{\rho} & \bar K-2 H_{\bar \psi \bar \psi}^{\rho}
\end{array}
\right),
\end{equation}
where $H_{\psi \psi}^{\rho}$, $H_{\psi \bar \psi}^{\rho}$, 
$H_{\bar \psi \psi}^{\rho}$ and $H_{\bar \psi \bar \psi}^{\rho}$ are
as defined in (\ref{hpsipsi2}), (\ref{hpsibpsi2}), (\ref{hbpsipsi2}) 
and (\ref{hbpsibpsi2}) respectively.

If $(\lambda,\phi)$ is an eigenpair of $J$, we can easily show that
\[
2\left(
\begin{array}{cc}
H_{\psi \psi}^{\rho}      &  H_{\psi \bar \psi}^{\rho} \\
H_{\bar \psi \psi}^{\rho} &  H_{\bar \psi \bar \psi}^{\rho}
\end{array}
\right)
\left(
\begin{array}{c}
\phi \\
\barphi
\end{array}
\right)
=( 1-\lambda)
\left(
\begin{array}{cc}
 K & 0 \\
0 & \bar{K}
\end{array}
\right)
\left(
\begin{array}{c}
\phi \\
\barphi
\end{array}
\right).
\]
If we again let $t_{ji}\equiv |t_{ji}|e^{i\mu_{ji}}$, 
$\zeta_{ji} \equiv |\zeta_{ji}|e^{i\theta_{ji}}$ and 
$\beta_{ji}$ be the $j$th component of the vectors 
$t_i = FQ_i\phi$, $z_i = FQ_i\psi$
and $b_i$ respectively, we can easily show that
\begin{eqnarray}
\lambda &=&  \frac{\sum_{i=1}^k\sum_{j=1}^n \sin^2(\mu_{ji}-\theta_{ji})|t_{ji}|^2 \beta_{ji}/|\zeta_{ji}| }
{\sum_{i=1}^k\sum_{j=1}^n |t_{ji}|^2}.
\label{jeigval}
\end{eqnarray}
Clearly, when $\beta_{ji} \leq |\zeta_{ji}|$ for all $j = 1,2,...,m$ and $i = 1,2,...n$,
$|\lambda| \leq 1$, and the fixed point iteration is guaranteed to converge
to at least a local minimizer of $\rho$.

The fixed point of $f$ may also be obtained by applying Newton's algorithm
to seek the root of $r(\psi) = 0$, where $r(\psi) = \psi - f(\psi)$. The Newton's 
method produces a sequences of iterates $\{\psi^{(\ell)}\}$ that satisfy
\[
\psi^{(\ell+1)} = \psi^{(\ell)} - J(\psi^{(\ell)})^{-1}r(\psi^{(\ell)}),
\]
where the $J$ matrix here is the Jacobian of $r$ with respect to $\psi$.
This approach is equivalent to applying Newton's algorithm (with appropriate 
line search and trust region strategies) to minimize $\rho(\psi)$.

Successive approximations to $J$ can be constructed from $\psi^{(\ell)}$
and $r(\psi^{(\ell)})$ using Broyden's technique. This is similar to
the Quasi-Newton algorithm discussed in the previous section.
As a special case, replacing $J$ with the crudest approximation, the 
identity matrix $I$, yields the standard error reduction algorithm.

If we multiply (\ref{fixp1}) from the left by $Q_i$ for 
$i = 1,2,...,k$,  and let $y^{(\ell)}=Q\psi^{(\ell)}$, 
where $Q = (Q_1^{\ast} \ \ Q_2^{\ast} \ \ ... \ \  Q_k^{\ast})^{\ast}$, we obtain
\begin{equation}
y^{(\ell+1)} = P_Q P_F (y^{(\ell)}),\, 
 \label{fixp3}
\end{equation}
where $P_Q = Q (Q^{\ast}Q)^{-1}Q^{\ast}$, and 
\[
P_F(y) = \hat{F}^{\ast} \frac{y}{|y|} \cdot b,
\]
where $\hat{F} = \Diag{F,F,...,F}$ and 
$b = (b_1^T \ \ b_2^T \ \ ... \ \ b_k^T)^T$.

Because a fixed point $y$ of $P_QP_F$ is in the range of $Q$, which is
typically full rank when $mk > n$, we may recover the corresponding fixed 
point of $f$ from $y$ via the least squares solution 
$\psi^{(\ell)}=(Q^{\ast}Q)^{-1}Q^{\ast}y^{(\ell)}$.

This nonlinear map is the composition of a (linear) orthogonal projector 
$P_Q$ and a (nonlinear) Fourier magnitude projector $P_F$. A fixed point 
iteration based on (\ref{fixp3}) is also called {\em alternating projection} 
(AP) algorithm in the phase retrieval literature because the approximation to the 
solution of (\ref{fixp3}) is obtained by applying $P_Q$ and $P_F$ in an 
alternating fashion.

It is easy to verify that $P_F$ is indeed a projection operator in the
sense that 
\begin{equation}
\|P_F(y) - y \| \leq \| w - y\| \ \ \mbox{for all} \ \ w \in \{w| w = P_F(w)\}.
\label{fproj}
\end{equation}
This property of $P_F$, together with the fact
that $P_Q$ is an orthogonal projection operator, i.e. $\| P_Q y - y \| \leq \| w - y\|$
for all $w \in \mbox{Range}(Q)$, allows us to show that the residual
error $\| P_QP_F(y^{(\ell)}) - y^{(\ell)}\|$ 
decreases monotonically in the AP algorithm. 
The proof of this observation was shown by Fienup in \cite{hio}, which 
we summarize below.

Let $\yell$ be the vector produced in the $\ell$-th AP iterate. 
Clearly, $\yell \in \mbox{Range}(Q)$.  Because $P_Q$ is an
orthogonal projector, we have
\begin{equation}
\| P_QP_F(\yell) - P_F(\yell) \| \leq \| P_QP_F(\yell) - \yell\|
=\|\yellu - \yell\|.
\label{pqy}
\end{equation}
Because $P_F(\yell) \in \{w | w = P_F(w)\}$, it follows from (\ref{fproj}) that
\begin{equation}
\| P_F(\yellu) - \yellu \| = \| P_F(P_QP_F(\yell)) - P_QP_F(\yell) \|
\leq \| P_QP_F(\yell) - P_F(\yell) \|.
\label{pfy}
\end{equation}
Consequently, we can deduce from (\ref{pqy}) and (\ref{pfy}) that
\[
\| P_F(\yellu) - \yellu \| \leq \| \yellu - \yell \|.
\]
Finally, it follows from the following inequality
\[
\| P_Q(P_F(\yellu) - \yellu) \| \leq \| P_F(\yellu) - \yellu \|,
\]
and the fact that $\yellu \in \mbox{Range}(Q)$ that
\begin{equation}
\| \yelld - \yellu \| \leq \| \yellu - \yell \|.
\label{apconv}
\end{equation}
The equality in (\ref{apconv}) holds only when $P_F(\yell) = \yell$, i.e., when 
convergence is reached. 

The inequality (\ref{apconv}) shows that the AP algorithm converges
to a stationary point. However, the convergence can be extremely
slow because
\[
\|z^{(\ell+1)}\| = \| F \yellu \| =\| \yellu \|
= \| P_Q P_F(\yell) \| \leq \| P_F(\yell) \| = \|b\|,
\]
and many of the terms $\beta_{ji}/\zeta_{ji}$, $i = 1,2,...,k$ and $j = 1,2,...,m$, in (\ref{jeigval}) may be great than 1. 
Only when $\yell$ is very close to the fixed point of 
$P_QP_F$, the spectral radius of the Jacobian of (\ref{fxpoint}) 
may become much smaller than 1 in (\ref{jeigval}) due to the reduction 
effect of the $\sin^2(\mu_{ji}-\theta_{ji})$ terms.

The simple alternating projection algorithm has been extended to 
the hybrid input-output (HIO) algorithm \cite{hio}, the relaxed averaged 
alternating reflection (RAAR) algorithm \cite{raar}, and many other variants 
\cite{saddle, Marchesini2007a} in the phase retrieval literature.  Just to give a few
examples, in the HIO and RAAR algorithms, the approximation to the
solutions of (\ref{fixp2}) and (\ref{fixp3}) are updated by
\begin{eqnarray*}
y^{(\ell+1)} &=& \left [ P_QP_F +
(I-P_Q)(I- \beta P_F)\right ] y^{(\ell)}\text{, HIO,} \\
y^{(\ell+1)} &=& \left [ 2 \beta P_Q P_F+(1-2 \beta) P_F +\beta (P_Q-I)\right ]
y^{(\ell)}\text{, RAAR.}\\
\psi^{(\ell+1)} &=& (Q^{\ast}Q)^{-1}Q^{\ast} y^{(\ell)},
\end{eqnarray*}

where $\beta$ is a constant often chosen to be between 0 and 1.

Although these algorithms tend to accelerate the convergence of 
$y^{(\ell)}$, their convergence behavior is less predictable 
and not well understood.

\section{Wigner Deconvolution} \label{sec:wigner}
Long before iterative methods were applied to solve the ptychography
problem, Rodenburg and his colleagues suggested that the problem
can be solved via what they called Wigner deconvolution \cite{bates-rodenburg}.

To explain the basic idea behind Wigner deconvolution, we need to 
state a continuum version of the ptychography problem.
If the set of translation vectors $\{\bx\}$ forms a continuum in 2D, then
it can be shown \cite{rodenburg} that the Fourier transform of 
$y_{\bx}^2 \equiv | \dft{a(\br)\hpsi(\br+\bx)}|^2$ with respect to
$\bx$, which we denote by $\dftx{y_{\bx}^2}$, can be written as the
convolution of two functions with respect to $\br'$, i.e.,
\begin{equation}
\dftx{y_{\bx}^2(\br)} = [A(\br')\bar{A}(\br'+\bx')] \star_{\br'}
[\Psi(\br')\bar{\Psi}(\br'-\bx')],
\label{wigner}
\end{equation}
where $A(\br') = \dft{a(\br)}$, $\Psi(\br')=\dft{\psi(\br)}$, $\bar{A}$ 
denotes the conjugate of $A$, $\bar{\Psi}$ denotes the conjugate of
$\Psi$, and $\star_{\br'}$ denotes a convolution operation with respect to ${\br'}$.
Note that $\dftx{y_{\bx}^2(\br)}$ is a function of $\bx'$. 
The Fourier transform of $A(\br')\bar{A}(\br'+\bx')$ or
$\Psi(\br')\bar{\Psi}(\br'-\bx')$ is called a {\em Wigner
distribution} in \cite{bates-rodenburg}.

The Fourier transforms used in the definition of $A(\br')$ and
$\Psi(\br')$ can be replaced by discrete Fourier transforms (DFT)
if both $a(\br)$ and $\psi(\br)$ are band-limited and
they are sampled at or beyond the Nyquist frequency. The Fourier 
transform of $y_{\bx}^2$ with respect to $\bx$ can be replaced by a 
DFT only if the translation vector $\bx$ is
sampled at or beyond the Nyquist frequency of $\psi(\br)$.  


We will define a fully sampled $\Psi(\br')$ by a column vector
\[
f = (f_1 \: f_2 \: \cdots f_n)^T,
\]
where $f_i = \Psi(\br'_i)$. Note that, when appeared by itself in $\Psi$, 
the variable $\bx'$ and $\br'$ can be used interchangeably, i.e., 
$f_i = \Psi(\bx'_i)$ holds also.

There are at least two ways to represent $\Psi(\br')\bar{\Psi}(\br'-\bx')$ 
systematically in a vector form.  We choose to write it as
\[
u(f) = 
\left(
\begin{array}{c}
\Diag{f} P_1^T \barf \\
\Diag{f} P_2^T \barf \\
\vdots \\
\Diag{f} P_n^T \barf
\end{array}
\right),
\]
where $P_i$ is a permutation matrix that shifts $\barf$ cyclically by
$i-1$ pixels, and $\barf$ denotes the conjugate of $f$.
This representation corresponds to writing down 
$\Psi(\br')\bar{\Psi}(\br'-\bx')$ by
having $\br'$ as the fastest changing index. By enumerating $\bx'$ first,
we can represent $\Psi(\br')\bar{\Psi}(\br'-\bx')$ in an alternative
form
\begin{equation}
\Pi u(f) = \left(
\begin{array}{c}
f_1 P_1^{T} \bar{f} \\
f_2 P_2^{T} \bar{f} \\
\vdots \\
f_n P_n^{T} \bar{f}
\end{array}
\right),
\label{piuf}
\end{equation}
where $\Pi$ is an $n^2 \times n^2$ permutation matrix that
reorders $\bx'$ and $\br'$.

Employing the same ordering we use to represent the fully 
sampled $\Psi(\br')\bar{\Psi}(\br'-\bx')$, we can express the
convolution kernel $A(\br')A(\br'+\bx')$ by a matrix $W$.
This matrix has a block diagonal form, i.e.,
\[
W = \left(
\begin{array}{cccc}
W_1 &     &        & \\
    & W_2 &        & \\
    &     & \ddots & \\
    &     &        & W_n 
\end{array}
\right),
\]
where $W_i$ is a block cyclic matrix with cyclic blocks (BCCB). 
This type of BCCB structure allow the convolution $W_i \Diag{f}P_i\barf$ to 
be carried out efficiently by using FFTs.

Using the notation established above, we can now express the 
sampled version of (\ref{wigner}) as
\[
\Pi W u(f) = \barb^2, 
\]
where 
\[
{\barb}^2 = \hat{F} \Pi
\left(
\begin{array}{c}
b_1^2 \\
b_2^2 \\
\vdots \\
b_m^2 
\end{array}
\right), 
\ \
\hat{F} = \left(
\begin{array}{cccc}
F &   &        & \\
  & F &        & \\
  &   & \ddots & \\
  &   &        & F 
\end{array}
\right),
\]
and $F$ is the matrix representation of a 2D discrete Fourier
transform of an image with $n$ pixels.  

If $W$ is nonsingular, i.e., $W_i$ is nonsingular for all $i = 1, 2,...,n$,
we can recover $u(f))$ by simply inverting $W$, i.e.
\begin{equation}
u(f) =  W^{-1} \Pi^T {\bar b}^2,
\label{modwig}
\end{equation}
Equation (\ref{modwig}) represents a deconvolution process, and 
is known as {\em Wigner deconvolution} \cite{bates-rodenburg}. 
The application of $W^{-1}$ to the vector $\Pi^T {\bar b}^2$ can be
achieved through an FFT based fast deconvolution or an iterative solver 
such as the conjugate gradient algorithm. We do not need to explicitly 
invert the matrix $W$.
If $W$ is singular or ill-conditioned, we may add a small constant to 
the diagonal of $W$ to regularize the deconvolution. 

Applying the permutation $\Pi$ to $u(f)$ allows us to rewrite the solution
of the deconvolution problem in the form of (\ref{piuf}).
If $f_i \neq 0$ for $i=1,2,...,n$, we define $c_i = 1/f_i$. 
Furthermore, let us define $\hat{g}^2 = \Pi W^{-1} \Pi^T {\bar f}^2$,
which can be partitioned as
\[
\hat{g}^2 = \left(
\begin{array}{c}
\hat{g}_1^2 \\
\hat{g}_2^2 \\
\vdots \\
\hat{g}_n^2
\end{array}
\right).
\]
where $\hat{g}_i^2 \in \mathbb{C}^{n \times 1}$.

By treating $c_i$ as a separate set of unknowns, with the exception of 
of $c_1$, which we will set to an arbitrary constant, e.g., 1, we can
turn (\ref{modwig}) into a linear least squares problem  by minimizing the
norm of
\begin{equation}
r =
\left(
\begin{array}{cccc}
P_1^T  & & & \\
P_2^T  & -\Diag{\hat{g}_2^2} &        & \\
\vdots &                    & \ddots  & \\
P_n^T  &                    &         & -\Diag{\hat{g}_n^2} \\
\end{array}
\right)
\left(
\begin{array}{c}
\hat{f} \\
c_2 \\
\vdots \\
c_n
\end{array}
\right)
-
\left(
\begin{array}{c}
c_1 \hat{g}_1^2 \\
0 \\
\vdots \\
0
\end{array}
\right).
\end{equation}
The minimization of $\|r\|$ can be easily solved by back 
substitution.  This is essentially the ``stepping out" procedure
described in \cite{bates-rodenburg}.  The reason that we can set $c_1$ to 
an arbitrary constant is that we are often interested in the relative amplitudes and phases of 
$\hat{\psi}(\br)$,
multiplying the entire image $\psi(\br)$ or $\Psi(\br')$ by a constant
does not change the quality of the image.

It may seem that the use of iterative method is not necessary 
if we can solve the ptychography problem by Wigner deconvolution,
which can be viewed as a linear inversion scheme.
However, as we will show below, the Wigner deconvolution 
problem cannot be solved directly (using an FFT based deconvolution
scheme) if $\bx$ is sampled below the Nyquist frequency, i.e. when
the amount of probe translation is larger than the resolution 
of the image to be reconstructed.

When $\bx$ is sampled below the Nyquist frequency, which can occur
in an experiment, we must modify (\ref{wigner}) by introducing an 
{\em aliasing operator} $S_{\bx'}$. 
Because $a(\br)$ is a localized window in practice, $A(\br')$ is 
subsampled in the reciprocal space. Therefore a subsampling operator 
$S_{\br'}$ must be included in a finite-dimensional
analog of (\ref{wigner}) to account for this effect.

With these additional operators, the sampled version of 
equation (\ref{wigner}) can be expressed as
\begin{equation}
S_{\bx'} \Pi S_{\br'} W u(f) = {\barb}^2,
\label{asub}
\end{equation}
where the dimensions of $\Pi$, $W$, $u(f)$ and ${\barb}^2$ need 
to be adjusted to reflect fewer pixel samples per diffraction frame
and fewer frames resulting from increased distance $\bx$ between two
adjacent frames. For simplicity, let us assume that $f$
and each frame $b_i^2$ are square images with $n$ and $m$ pixels 
respectively, and the distance between two adjacent frames is $d_x$
(in either the horizontal or the vertical direction).  Then,
the aliasing operator $S_{\bx'}$ in (\ref{asub}) is a block diagonal 
matrix consisting of $n_f$ diagonal blocks of dimension $m \times n$,
where $n_f = \lfloor \sqrt{n}/\sqrt{m} \rfloor$.  The 
subsampling operator $S_{\br'}$ is a block diagonal matrix
consisting of $m$ diagonal blocks of dimension $n_f \times n$,  and 
$\Pi$ is an $n_f m \times n_f m$ row permutation matrix
that reshuffles the rows of $S_{\br'} W u(f)$ so that $\bx'$ is
the fastest changing index. For 1D signals, a diagonal block of 
$S_{\bx'}$ can be represented by
\[
(I_m \ \ I_m \ \ \cdots \ \ I_m),
\]
where $I_m$ is an $m \times m$ identity matrix. Similarly, a typical 
diagonal block of $S_{\br'}$ has the form
\[
(\cdots \ \ 0 \ \ I_{n_f} \ \ 0 \ \ \cdots),
\]
where $I_{n_f}$ is an $n_f \times n_f$ identity matrix.

Because $S_{\bx'}$, $\Pi$ and $S_{\br'}$ are not 
square matrices, we cannot obtain $u(f)$ by simply applying 
the inverse of these matrices and $W^{-1}$ to 
${\barb}^2$.  

Instead, we must recover $f$, hence the fully sampled 
$\hat{\psi}(r)$, by solving the following nonlinear least 
squares problem
\begin{equation}
\min_{f} \| S_{\bx'} \Pi S_{\br'} W u(f) - {\barb}^2 \|^2.
\label{wigls}
\end{equation}

It is not difficult to see that the objective function in the nonlinear
least squares problem (\ref{wigls}) is equivalent to (\ref{obj2}).
Therefore, iterative optimization techniques applied to minimize 
(\ref{obj2}) can be used to solve (\ref{wigls}) also. However,
the evaluation of the objective function in (\ref{wigls}) and its
derivatives, which we will not show here, are more costly because 
evaluating $u(f)$ requires at least $\mathcal{O}(n^2)$ operations, 
and multiplying $W$ with $u(f)$ requires an additional 
$\mathcal{O}(n^2 \log(n))$ operations.  This operation count is
much higher than that associated with evaluating (\ref{obj2}), which
is $\mathcal{O}(mn_f\log(n_f) + mn_f)$.


We should mention that, if one is interested a reconstruction 
of limited resolution, $\tilde{f}$, which is a cropped version of 
$\Psi(-\bx')$, the objective function in (\ref{wigls}) can be modified 
to become
\[
\| S_{\bx'} \Pi (S_{\br'} W S_{\br'}^T) \tilde{u}(\tilde{f}) - 
{\barb}^2 \|^2,
\]
where $\tilde{u}(\tilde{f}) \in \mathbb{C}^{m \times 1}$.
Furthermore, if the translation of the frame $\bx'$ is chosen
to be commensurate with the size of each frame, e.g., $x' = \sqrt{n/m}$, 
then $S_{\bx'}$ becomes an identity matrix. Consequently, one may obtain 
$\tilde{u}(\tilde f)$ (and subsequently $\tilde{f}$) by performing a Wigner 
deconvolution. 


\section{Numerical Examples} \label{sec:example}
\begin{figure}[hbtp]
  \begin{center}
    \includegraphics[width=0.6\textwidth]{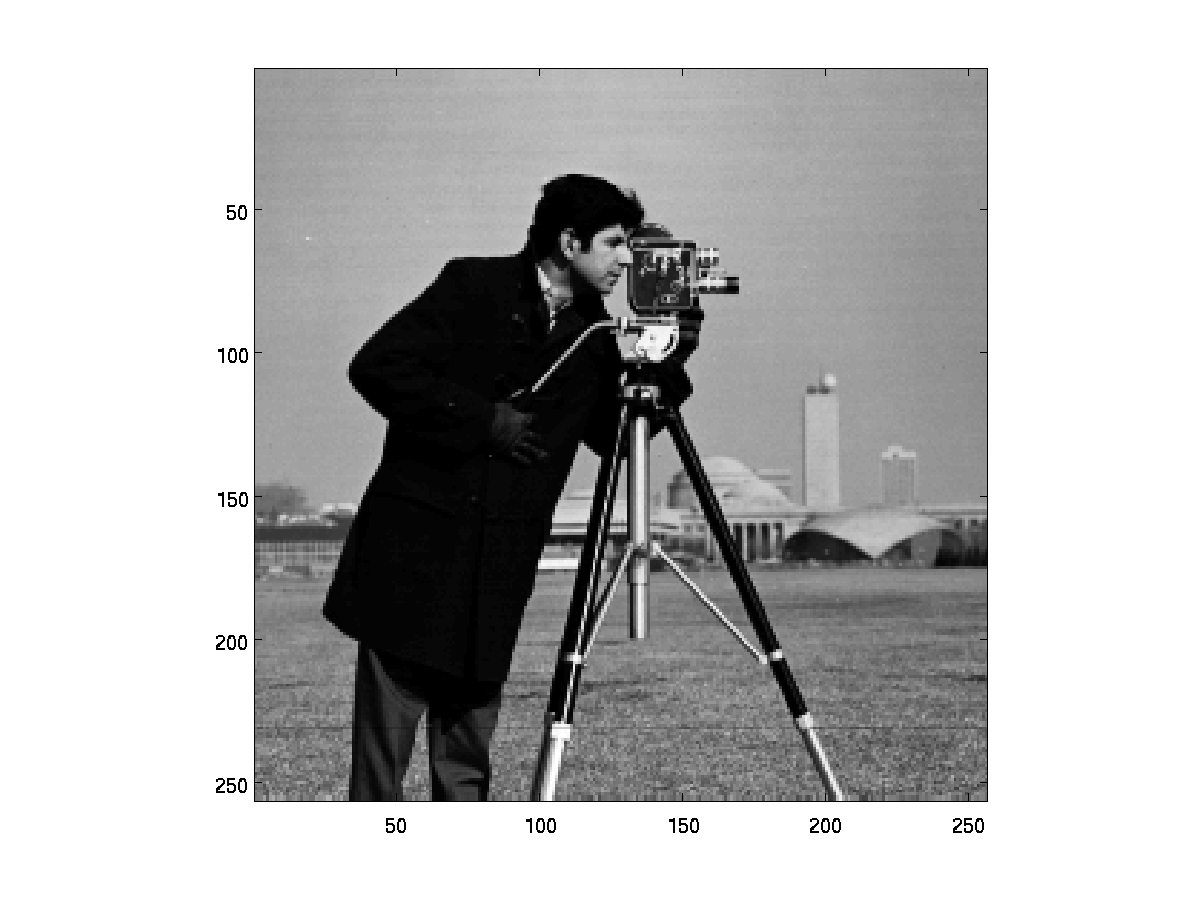}
  \end{center}
  \caption{The cameraman test image.}
  \label{fig:caman}
\end{figure}

\begin{figure}[hbtp]
\begin{center}
\hfill
\subfigure[Amplitude]{
    \includegraphics[width=0.45\textwidth]{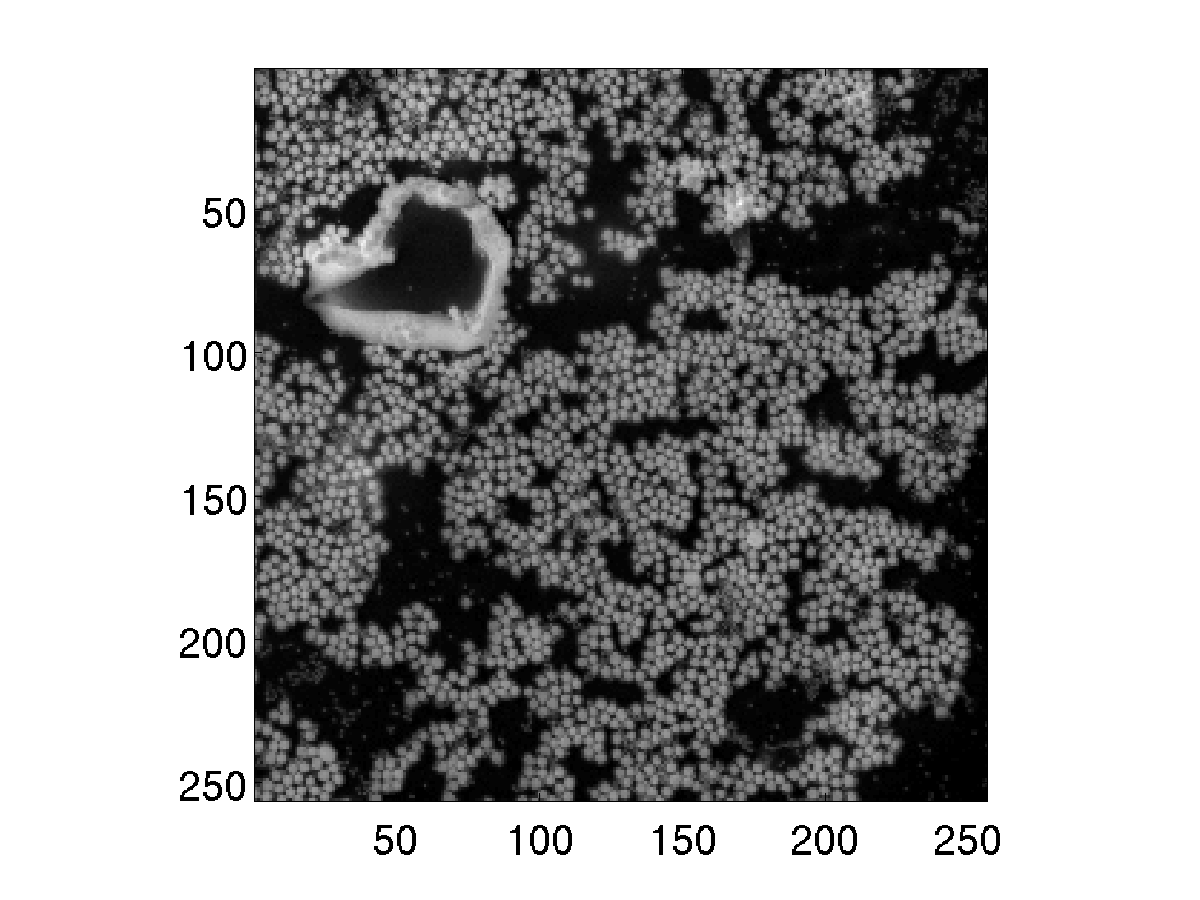}
    \label{fig:auamp}
}
\hfill
\hspace{0.25in}
\subfigure[Phase]{
    \includegraphics[width=0.45\textwidth]{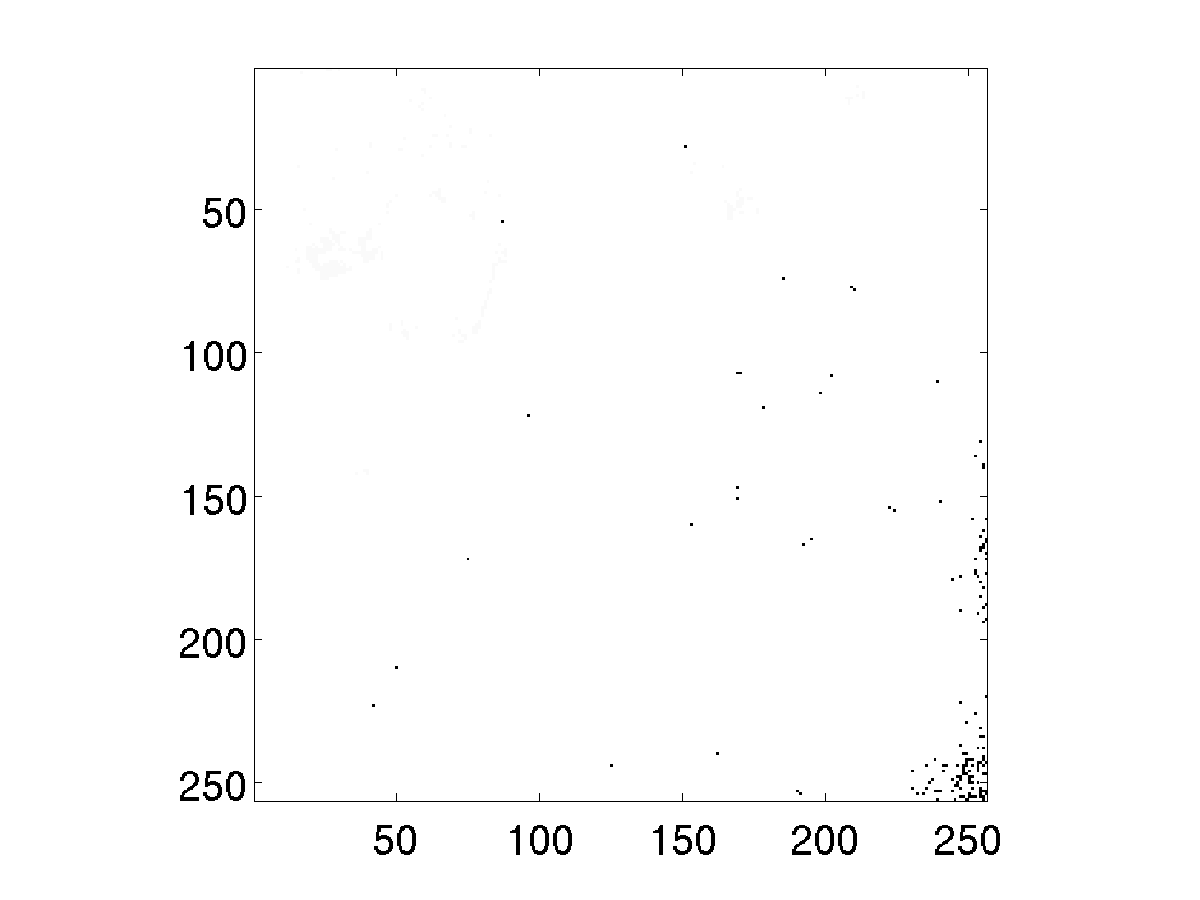}
    \label{fig:auphase}
}
\hfill
\end{center}
\caption{The amplitude and phase of the transmission coefficient of
         a collection of gold balls.}
\label{fig:auballs}
\end{figure}

In this section, we demonstrate and compare the convergence of 
iterative algorithms for ptychographic reconstruction using two test images.
The first test image is a $256 \times 256$ real-valued 
cameraman image shown in Figure~\ref{fig:caman}.  The image is often used 
in the image processing community to test image reconstruction and 
restoration algorithms.  The second test image is a complex valued image. 
It also contains $256\times 256$ pixels that correspond to the complex 
transmission coefficients of a collection of gold balls embedded in some 
medium.  The amplitude and phase angles of these pixels are shown in 
Figure~\ref{fig:auballs}.

All numerical examples presented in this paper are performed in 
MATLAB. 
\subsection{Comparison of Convergence Rate}
\begin{figure}[hbtp]
\begin{center}
\hfill
\subfigure[The binary probe used in the reconstruction of the cameraman image.]{
    \includegraphics[width=0.45\textwidth]{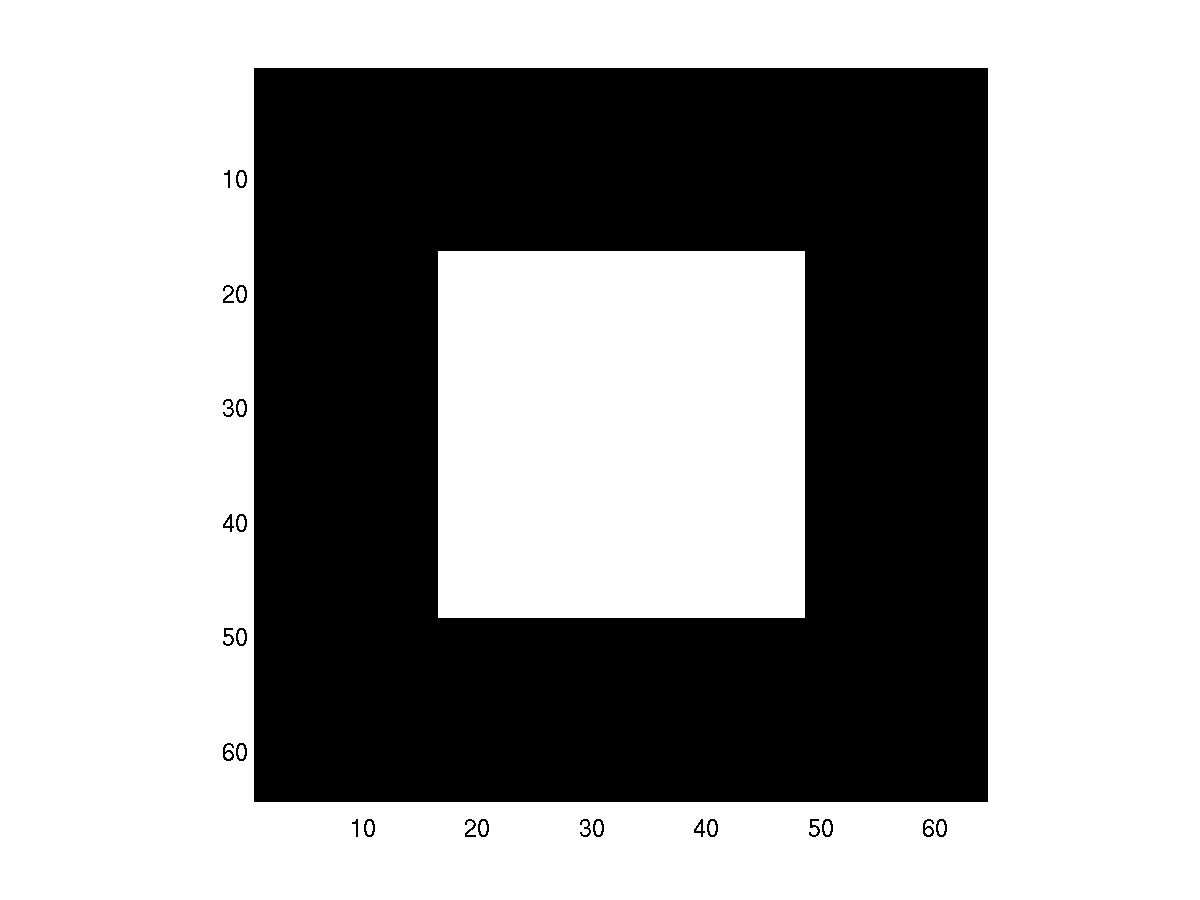}
    \label{fig:maskcam}
}
\hfill
\hspace{0.25in}
\subfigure[The probe used in the reconstruction of the gold ball image.]{
    \includegraphics[width=0.45\textwidth]{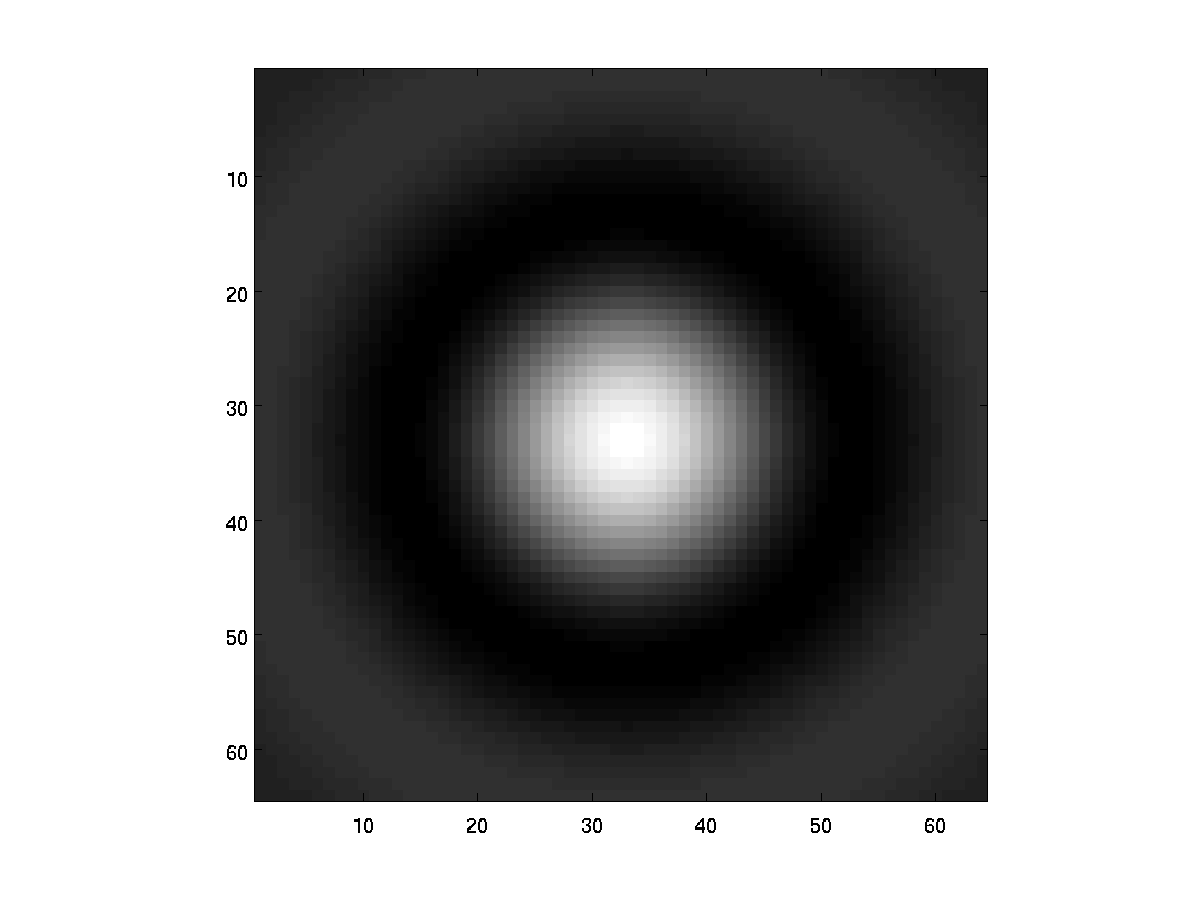}
    \label{fig:maskau}
}
\hfill
\end{center}
\caption{The illuminating probes $a(\br)$ used in ptychographic 
         reconstructions of the cameraman and gold ball images.}
\label{fig:aumask}
\end{figure}

In this section, we show the convergence behavior of different iterative 
algorithms we discussed in section~\ref{sec:algs} by numerical 
experiments.  In the cameraman image reconstruction experiment, we
choose the illuminating probe $a(\br)$ to be a $64 \times 64$ binary
probe shown in Figure~\ref{fig:maskcam}. 
The pixels within the $32\times 32$ square 
at the center of the probe assume the value of 1. 
All other pixels take the value of 0. The zero padding of the inner
$32\times 32$ square ensures that the diffraction pattern of a 
$64 \times 64$ frame associated with this probe is oversampled in 
the reciprocal space.  
In the gold ball image reconstruction experiment,
the illuminating probe is chosen to be the amplitude of the Fourier
transform of an annular ring with inner radius of $r_1 \approx 5.4$ and outer 
radius of $r_2 \approx 19.4$. This probe mimics the true illumination
used in a physical experiment.

In the cameraman experiment, the probe is translated by $8$ pixels at 
a time in either horizontal or vertical direction.
To prepare a stack of $k$ diffraction images $b_i$, $i = 1,2,...,k$,
we start from the upper left corner of the true image, extract a $64\times 64$ 
frame, and multiply it with the probe, and then apply a 2D FFT 
to the product. The magnitude of transform is recorded and saved
before we move either horizontally or vertically to obtain the next frame.
If the lower right corner of the frame goes outside of the image (which
does not happen in this particular case), we simply 
``wrap the probe around" the image as if the image is periodically
extended.  As a result, the total number of diffraction frames we
use for each reconstruction is
\[
k = \frac{256}{8} \cdot \frac{256}{8} = 1024.
\]

As we will show in section~\ref{sec:trans}, the size of translation, 
which determines the amount of overlap between adjacent frames,
has a noticeable effect on the convergence of the iterative 
reconstruction algorithms.

\begin{figure}[hbtp]
\centering
\hfill
\subfigure[Change of the relative residual norm (res) for the 
           reconstruction of the cameraman image.]{
    \includegraphics[width=0.45\textwidth]{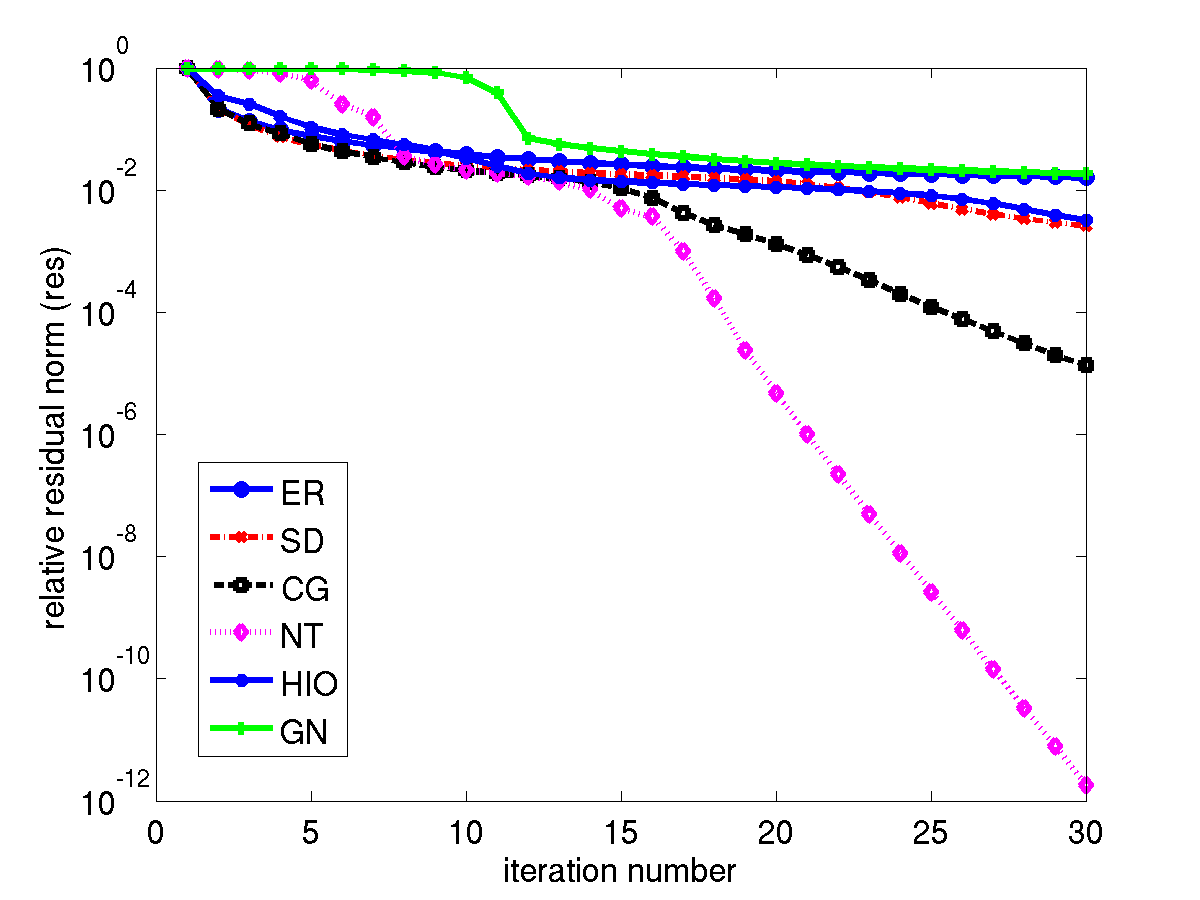}
    \label{fig:camres}
}
\hfill
\subfigure[Change of the relative error (err) for the reconstruction
           of the cameraman image.]{
    \includegraphics[width=0.45\textwidth]{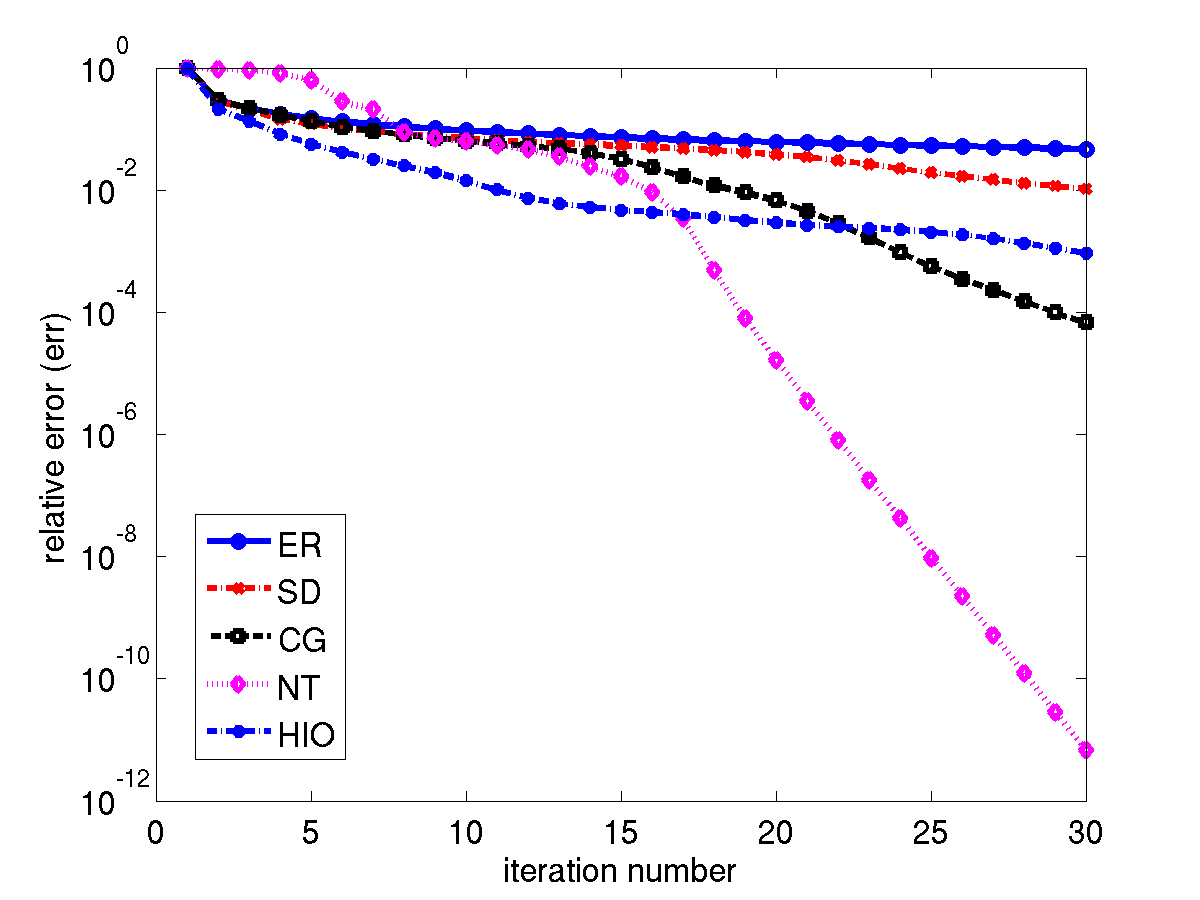}
    \label{fig:camerr}
}
\caption{A comparison of the convergence behavior of different 
         iterative ptychographic reconstruction algorithms for 
         the cameraman image.}
\label{fig:camconv}
\end{figure}

\begin{figure}[htbp]
\centering
  \hfill
  \subfigure[ER reconstruction]{
    \label{fig:cambyer}
    \includegraphics[width=0.45\textwidth]{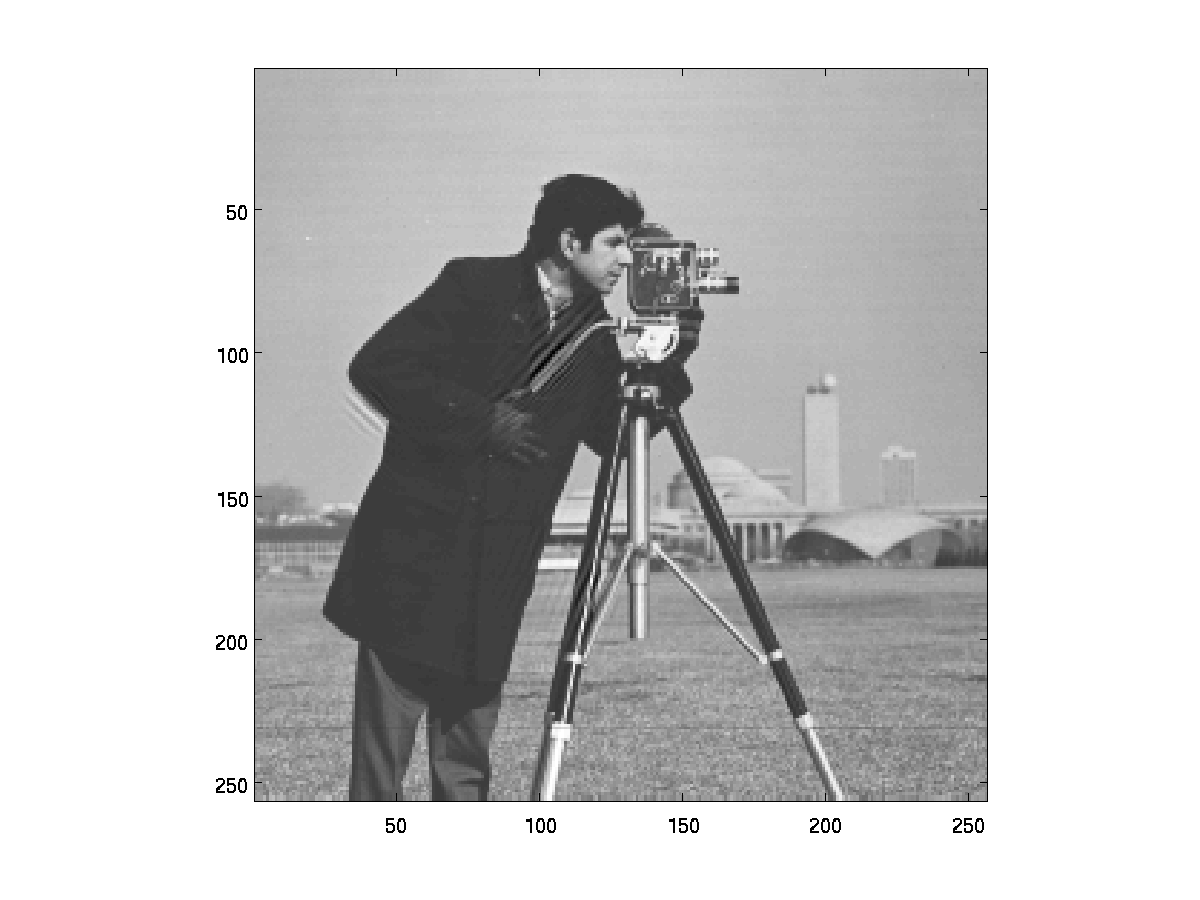}
  }
  \hfill
  \subfigure[SD reconstruction]{
    \label{fig:cambysd}
    \includegraphics[width=0.45\textwidth]{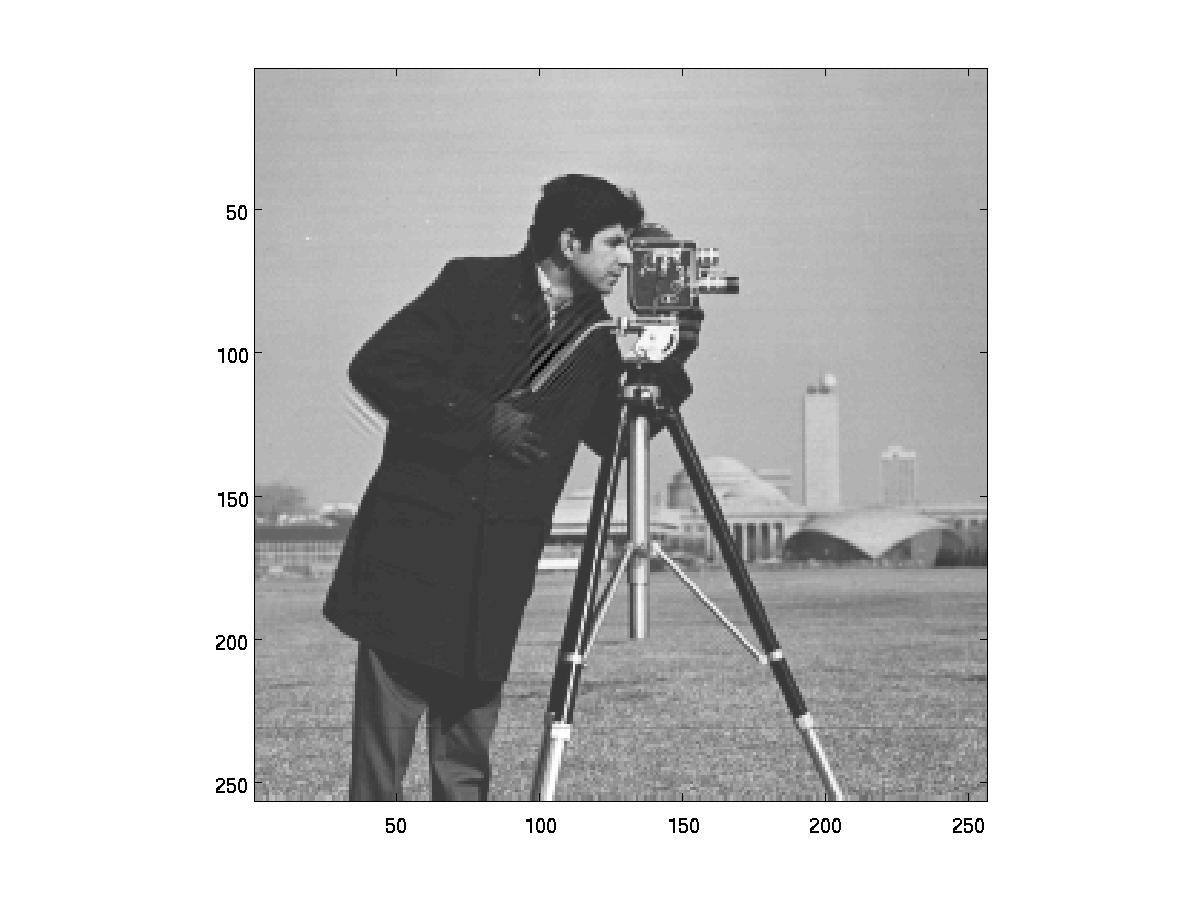}
  }
\caption{The reconstructed cameraman images by ER and SD algorithms contain
         visible ringing artifacts.}
\label{fig:camrecon}
\end{figure}

Figure~\ref{fig:camconv} shows the convergence history of several
iterative algorithms discussed in section~\ref{sec:algs} when they
are applied to the diffraction frames extracted from the cameraman
image.  
We plot both the relative residual norm defined by
\begin{equation}
res = \frac{\sqrt{\sum_{i=1}^k \| |z_i|^{(\ell)} - b_i \|^2}}
{\sqrt{\sum_{i=1}^k \| b_i\|^2}},
\label{relres}
\end{equation}
where $|z_i|^{(\ell)} = |FQ_i \psi^{(\ell)}|$ and $\ell$ is 
the iteration number, and the relative error of the reconstructed 
image defined by
\[
err = \frac{\| \psi^{(\ell)}-\hpsi \| } {\|\hpsi\|}.
\]
In these runs, an exact line search is used in the steepest descent (SD), 
nonlinear conjugate gradient (CG). The Steihaug's trust region technique
implemented in \cite{tkelley} is used in the Newton's method (NT).  
We set the starting guess of the solution $\hpsi$ to 
\[
\psi^{(0)} = \left(\sum_{i=1}^k Q_i^{\ast}Q_i\right)^{-1}
\sum_{i=1}^k Q_i^{\ast} b_i.
\]
It is clear from Figure~\ref{fig:camconv} that NT converges much faster 
than the other algorithms.  Its performance is followed by the CG algorithm 
which is much faster than the error reduction (ER), SD, Gauss-Newton (GN) 
and the hybrid input-output (HIO) algorithms.  Similar
convergence behavior is observed when other random starting guesses are used,
although occasionally, a random starting guess can lead to stagnation 
or convergence to a local minimizer. We will discuss this issue in 
section~\ref{sec:localmin}.  We set the maximum number of iterations 
allowed in all runs to 30. This is somewhat excessive for both NT and CG 
algorithms. Typically, when the relative error of the reconstructed image
falls below $10^{-3}$, it is nearly impossible to visually distinguish
the reconstruction from the true image.  When the relative error is larger,
the reconstructed cameraman images may contain visible artifacts 
such as those shown in Figures~\ref{fig:cambyer} and~\ref{fig:cambysd} 
which are produced at the end of the 30th ER and SD iterations respectively.

It is somewhat surprising that GN performs poorly on this problem.
We believe the problem is that we used the MATLAB implementation of
the large-scale Gauss-Newton algorithm, i.e., the function 
{\tt lsqnonlin} in the MATLAB's Optimization Toolbox, which does not
handle functions of complex variable very well.  Moreover, it is 
not easy to obtain the relative error associated with the approximate
reconstruction produced at each iteration from this function.

For the reconstruction of the gold ball image, we choose the
starting guess to be
\[
\psi^{(0)} = \left(\sum_{i=1}^k Q_i^{\ast}Q_i\right)^{-1}
\sum_{i=1}^k Q_i^{\ast} \Diag{b_i} \Diag{|u_i|}^{-1} u_i,
\]
where $u_i$ is a complex random vector, and the real and imaginary
part of each component has a uniform distribution within $[-1,1]$.

\begin{figure}[hbtp]
\centering
\hfill
\subfigure[Change of the relative residual norm (res) for the 
           reconstruction of the gold ball image.]{
    \includegraphics[width=0.45\textwidth]{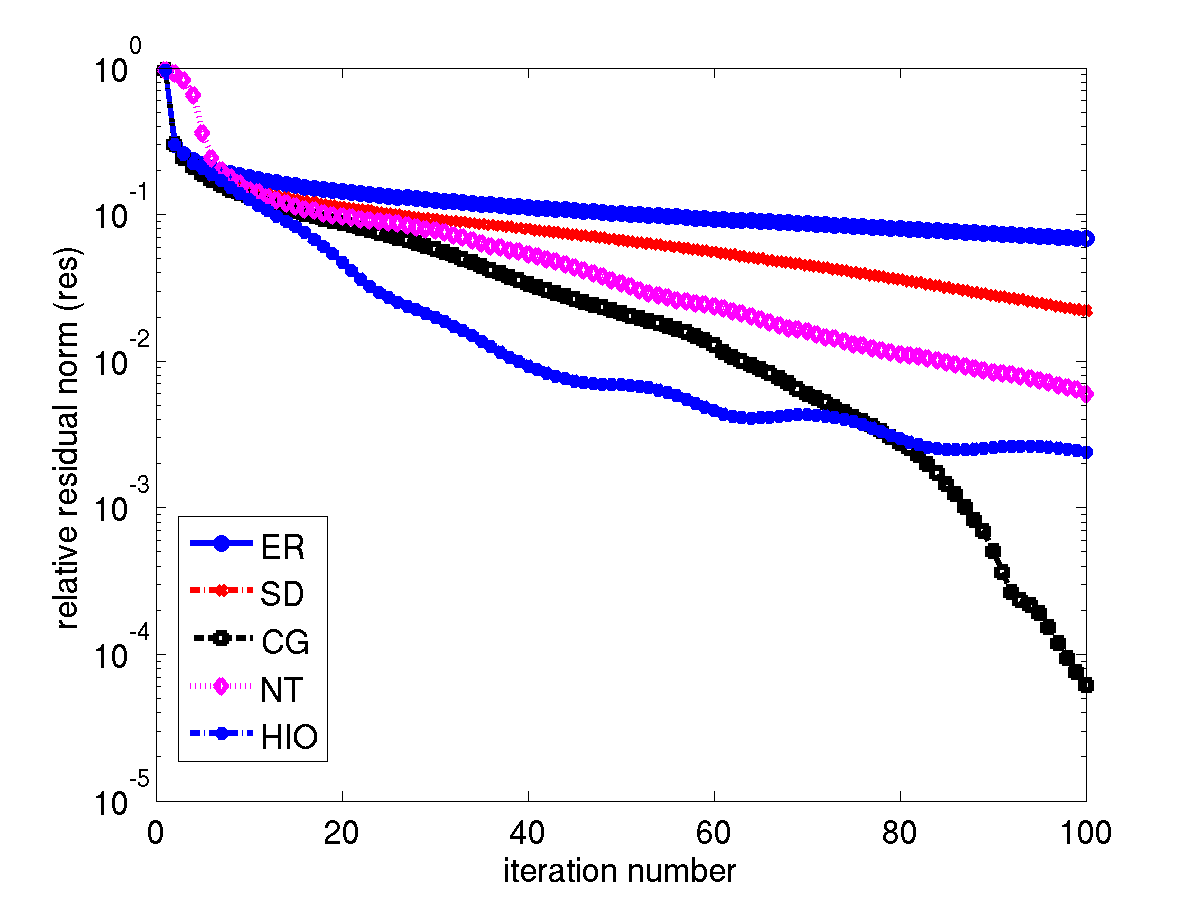}
    \label{fig:aures}
}
\hfill
\subfigure[Change of the relative error (err) for the 
           reconstruction of the gold ball image.]{
    \includegraphics[width=0.45\textwidth]{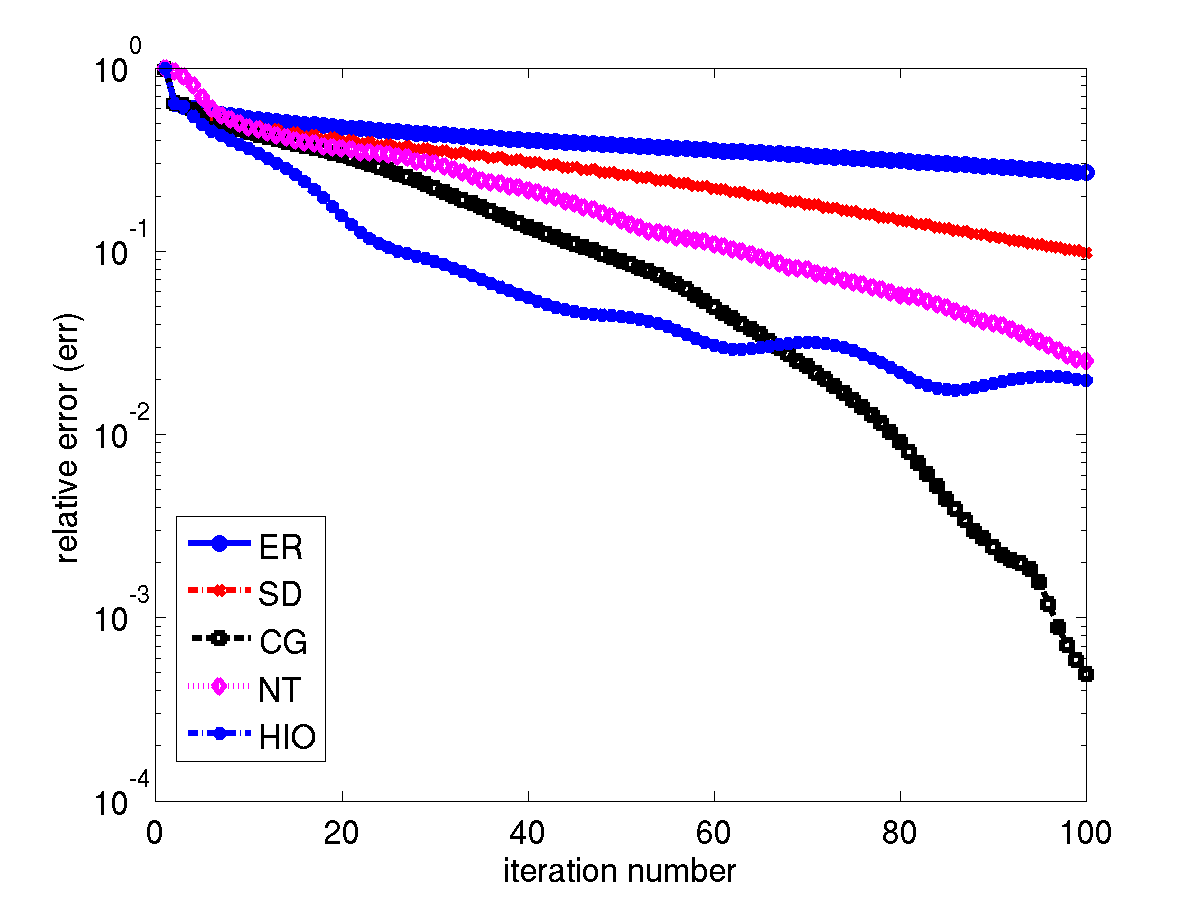}
    \label{fig:auerr}
}
\caption{A comparison of the convergence behavior of different 
         iterative ptychographic reconstruction algorithms for 
         the gold ball image.}
\label{fig:auconv}
\end{figure}

In this experiment, the probe is translated by a larger amount (16 pixels)
in either horizontal or vertical direction.
Figure~\ref{fig:auconv} shows the convergence history of ER, SD, CG, HIO,
and NT.  From Figure~\ref{fig:aures}, it appears that CG is the best 
among all the methods we tried. The HIO algorithm performs well 
in the first 60 iterations, but then stagnates. As we can see from
Figure~\ref{fig:auconv} that the neither the residual norm nor the 
relative error associated with HIO changes monotonically.  
This is not completely surprising because HIO does not try to minimize 
either objective functions.  For this example, the performance of NT 
lags behind CG by a large margin although both algorithms
exhibit monotonic convergence with a more predictable error reduction.
We should mention that to measure the relative error associated with a 
reconstructed gold ball image $\psi^{(\ell)}$, we need to multiply it 
by a constant phase factor $\gamma$ first, i.e., the relative error is 
defined as
\[
err = \frac{\|\gamma \psi^{(\ell)} -\hat{\psi} \| }{\|\hat{\psi}\|}.
\]

\begin{figure}[htbp]
\centering
  \hfill
  \subfigure[The magnitude of the reconstructed gold ball image
             produced by the CG algorithm.]{
    \label{fig:aumag_cg}
    \includegraphics[width=0.45\textwidth]{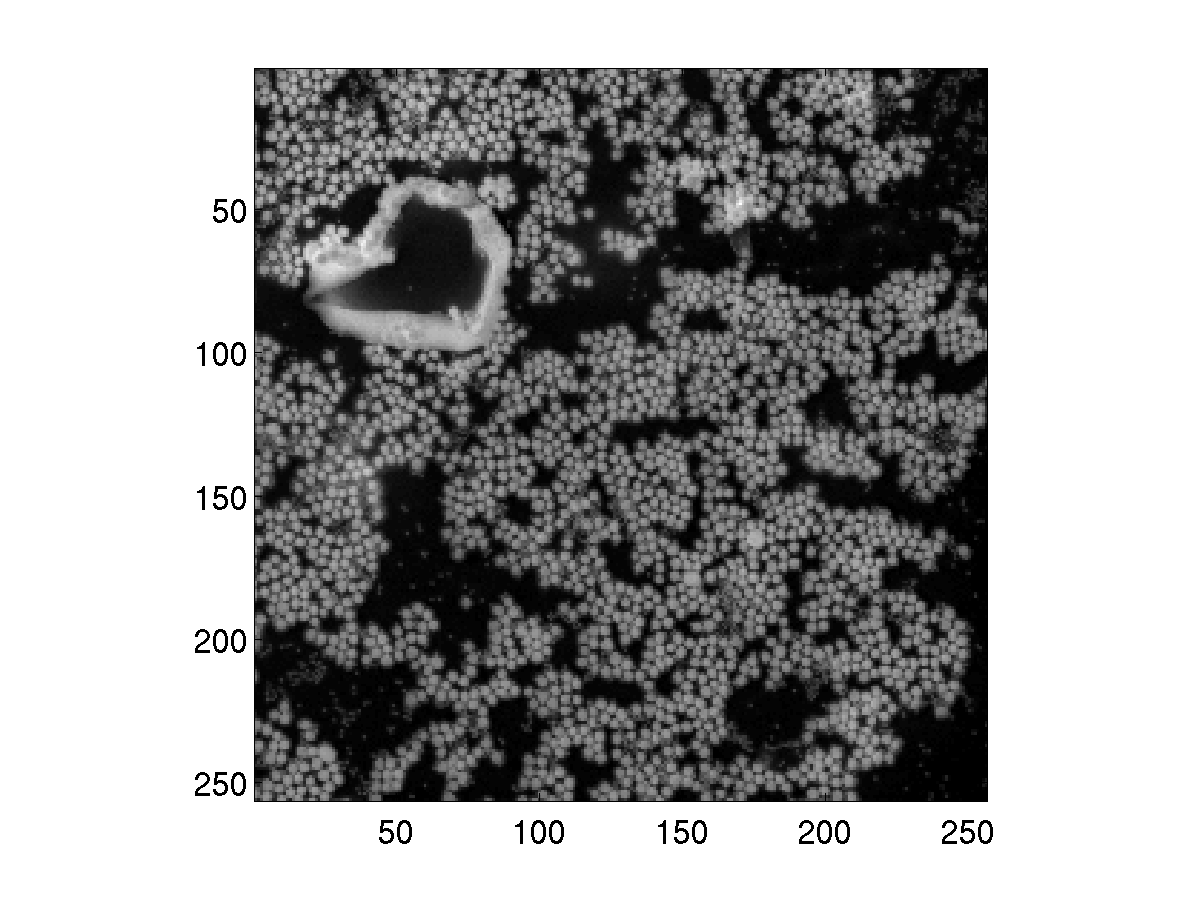}
  }
  \hfill
  \subfigure[The magnitude of the error associated with the 
             reconstructed gold ball image produced by the CG algorithm.]{
    \label{fig:auface_cg}
    \includegraphics[width=0.45\textwidth]{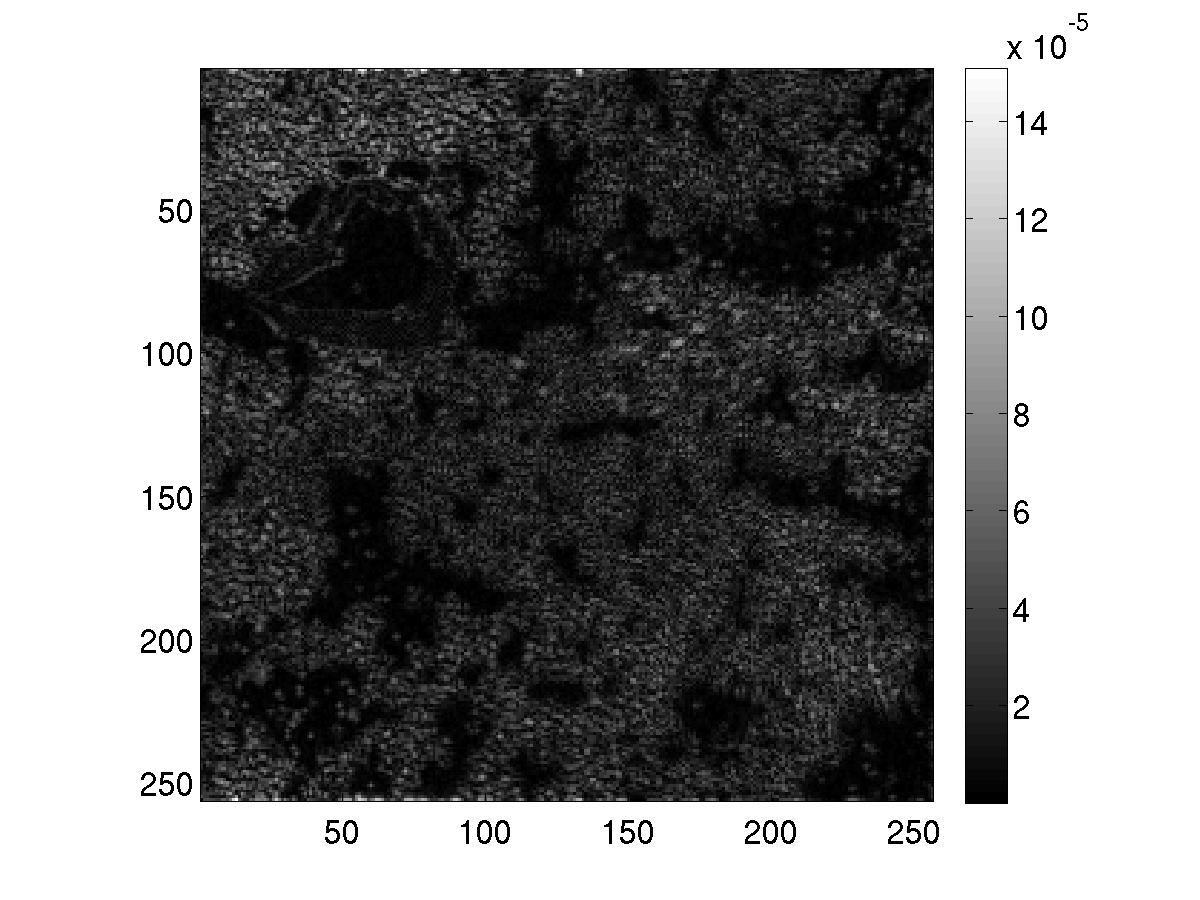}
  }
\\
  \hfill
  \subfigure[The magnitude of the reconstructed gold ball image
             produced by the HIO algorithm.]{
    \label{fig:aumag_hio}
    \includegraphics[width=0.45\textwidth]{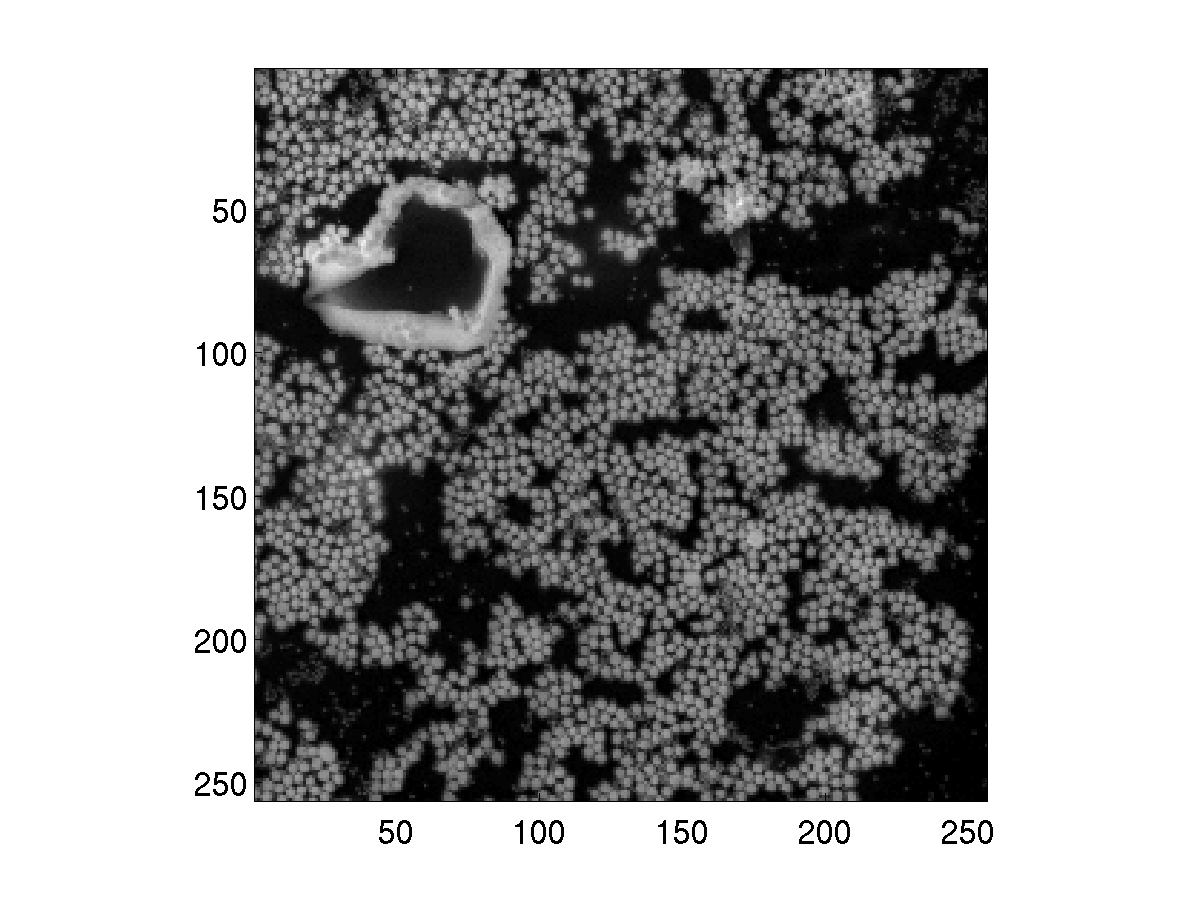}
  }
  \hfill
  \subfigure[The magnitude of the error associated with the 
             reconstructed gold ball image produced by the HIO algorithm.]{
    \label{fig:auface_hio}
    \includegraphics[width=0.45\textwidth]{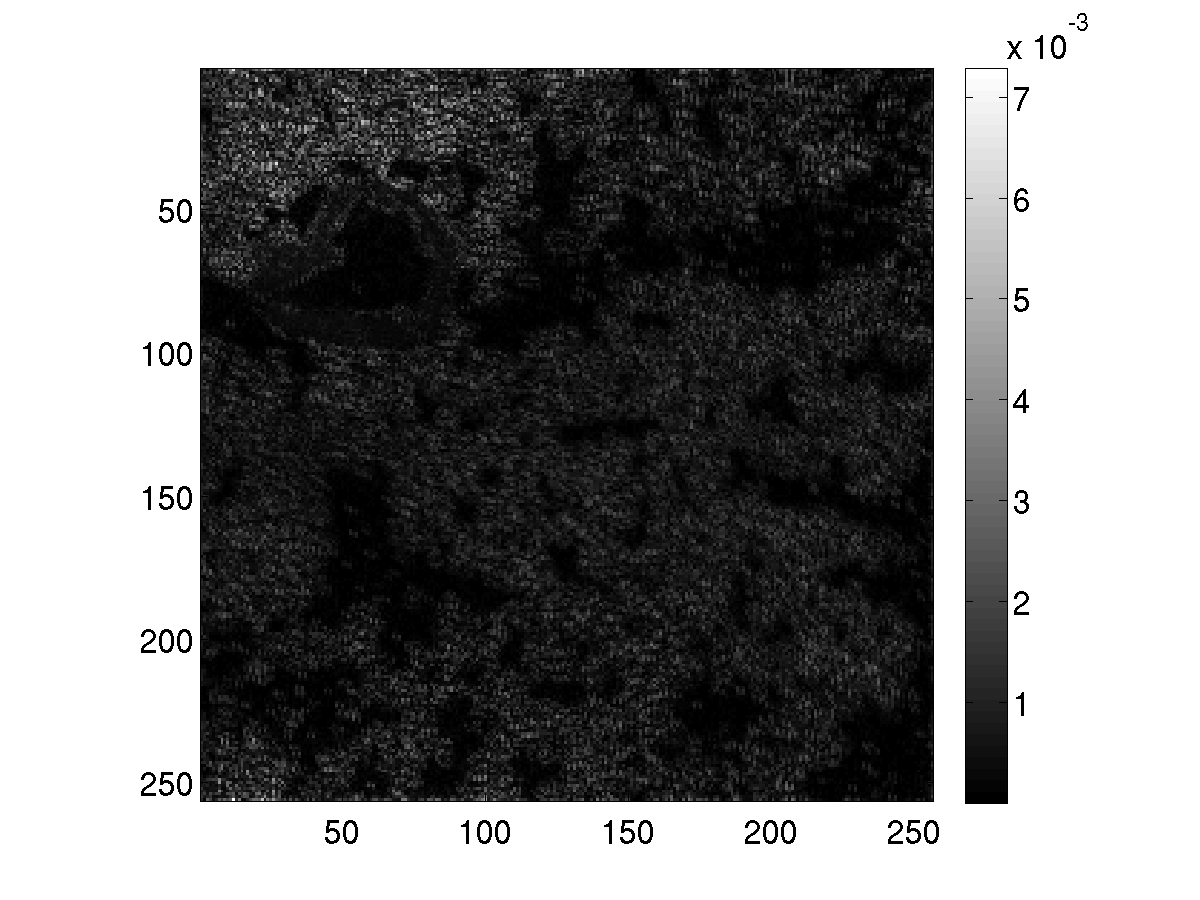}
  }
\caption{The reconstructed cameraman images produced by CG and HIO.}
\label{fig:aurecon}
\end{figure}

In Figure~\ref{fig:aurecon}, we can clearly see that 
the magnitude of the reconstructed images produced by
CG (Figure~\ref{fig:aumag_cg}) and HIO (Figure~\ref{fig:aumag_hio}) 
are nearly indistinguishable from the magnitude of the true image.
However, the phase angles of the reconstructed image produced
by CG (Figure~\ref{fig:auface_hio}) appear to be better than those
produced by HIO, which is indicated by the magnitude of the
absolute errors $|\gamma \psi^{(\ell)} -\hat{\psi}|$ shown in 
Figures~\ref{fig:auface_cg} and \ref{fig:auface_hio}.

\subsection{The Effect of Preconditioning}
\begin{figure}[hbtp]
\centering
\hfill
\subfigure[The effect of the preconditioner on the convergence of SD.]{
    \includegraphics[width=0.45\textwidth]{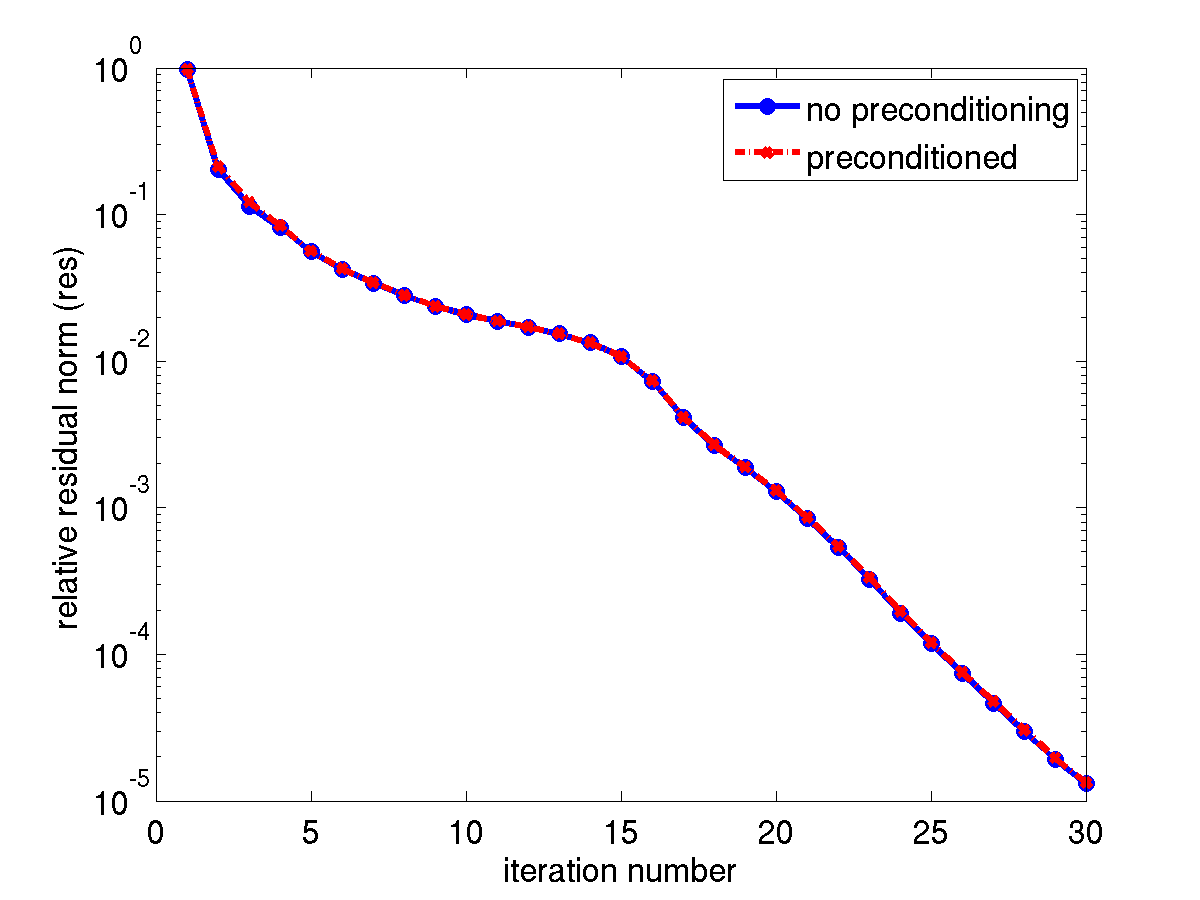}
    \label{fig:camcgprec}
}
\hfill
\subfigure[The effect of the preconditioner on the convergence of CG.]{
    \includegraphics[width=0.45\textwidth]{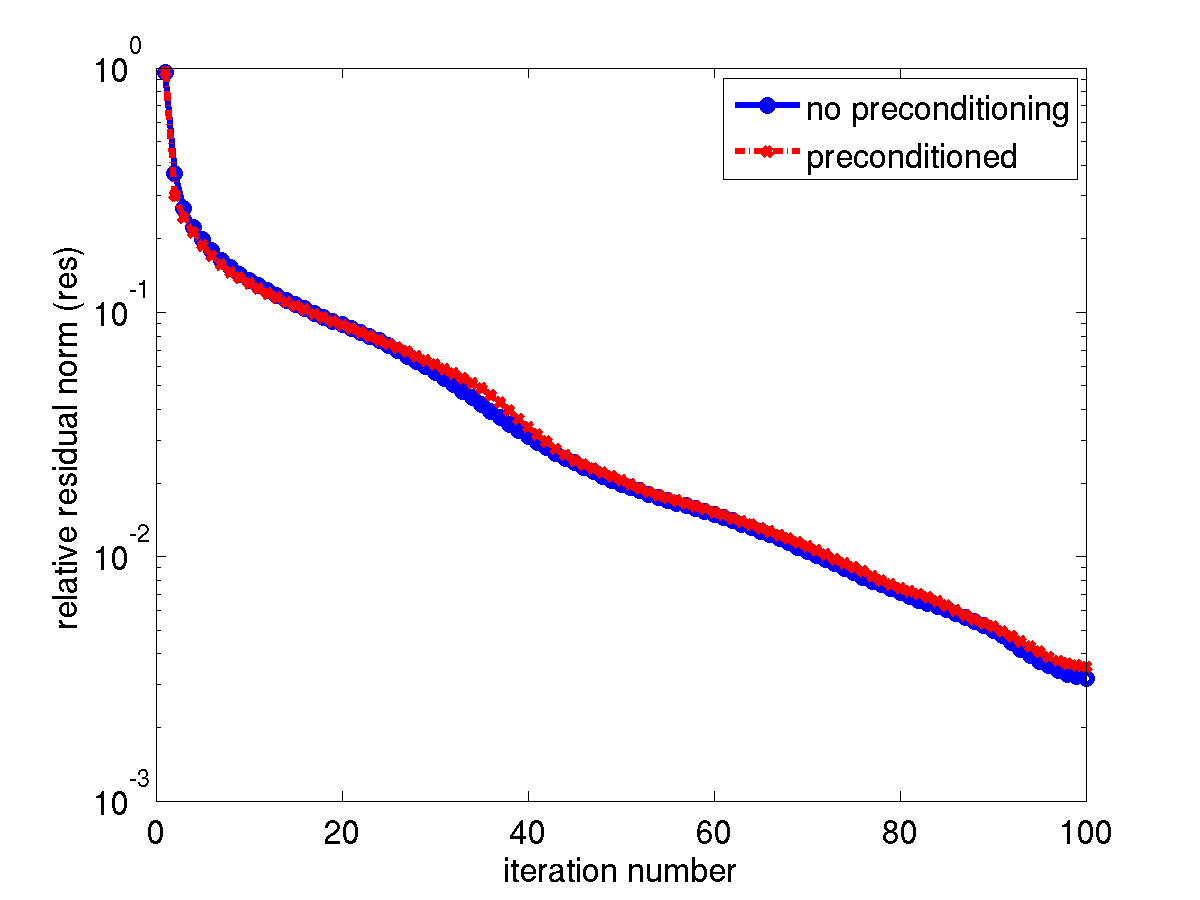}
    \label{fig:aucgprec}
}
\caption{The effect of a preconditioner on the convergence of
         the CG algorithms applied to cameraman and gold ball image 
         reconstruction.}
\label{fig:cgprec}
\end{figure}

As we indicated in Section~\ref{sec:prec}, the use of a preconditioner
can enhance the convergence of SD and CG.  A natural preconditioner 
that is easy to construct is (\ref{precond}).  However, this preconditioner
is only effective, when the condition number of $K$ is relatively large.
For the binary probe used in the reconstruction of the cameraman image,
$K = 4 I$. The condition number of this matrix is 1. Hence, using this
preconditioner has no effect on the convergence of the CG iteration, as
we can clearly see in Figure~\ref{fig:camcgprec}.  The condition number
associated with the probe used in the gold ball image reconstruction
is around 4.5. Hence the effect of the preconditioner is negligible as
we can see from Figure~\ref{fig:aucgprec}.

\subsection{Local Minimizer and the Choice of the Objective Function}
\label{sec:localmin}
As we indicated in section~\ref{sec:hess}, based on the analytic 
Hessian and curvature expression, that neither $\epsilon(\psi)$ 
nor $\rho(\psi)$ is globally convex.  This observation suggests
that all iterative optimization algorithm discussed above may
converge to a local minimizer.  Although we found that in practice,
local minimizers are not easy to find, they do exist as the following
example show. 

\begin{figure}[hbtp]
\centering
\hfill
\subfigure[True image.]{
    \includegraphics[width=0.45\textwidth]{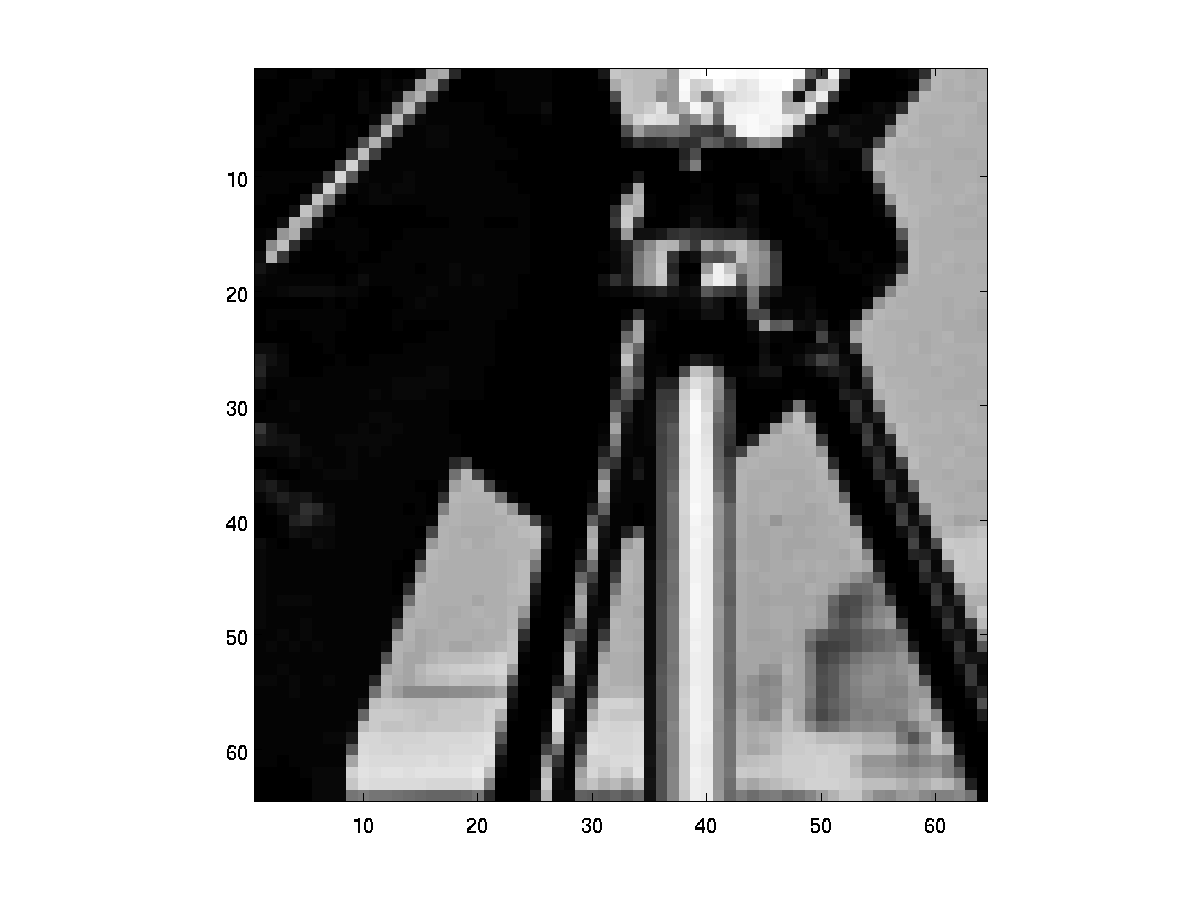}
    \label{fig:subcaman}
}
\hfill
\subfigure[The reconstructed image (a local minimizer).]{
    \includegraphics[width=0.45\textwidth]{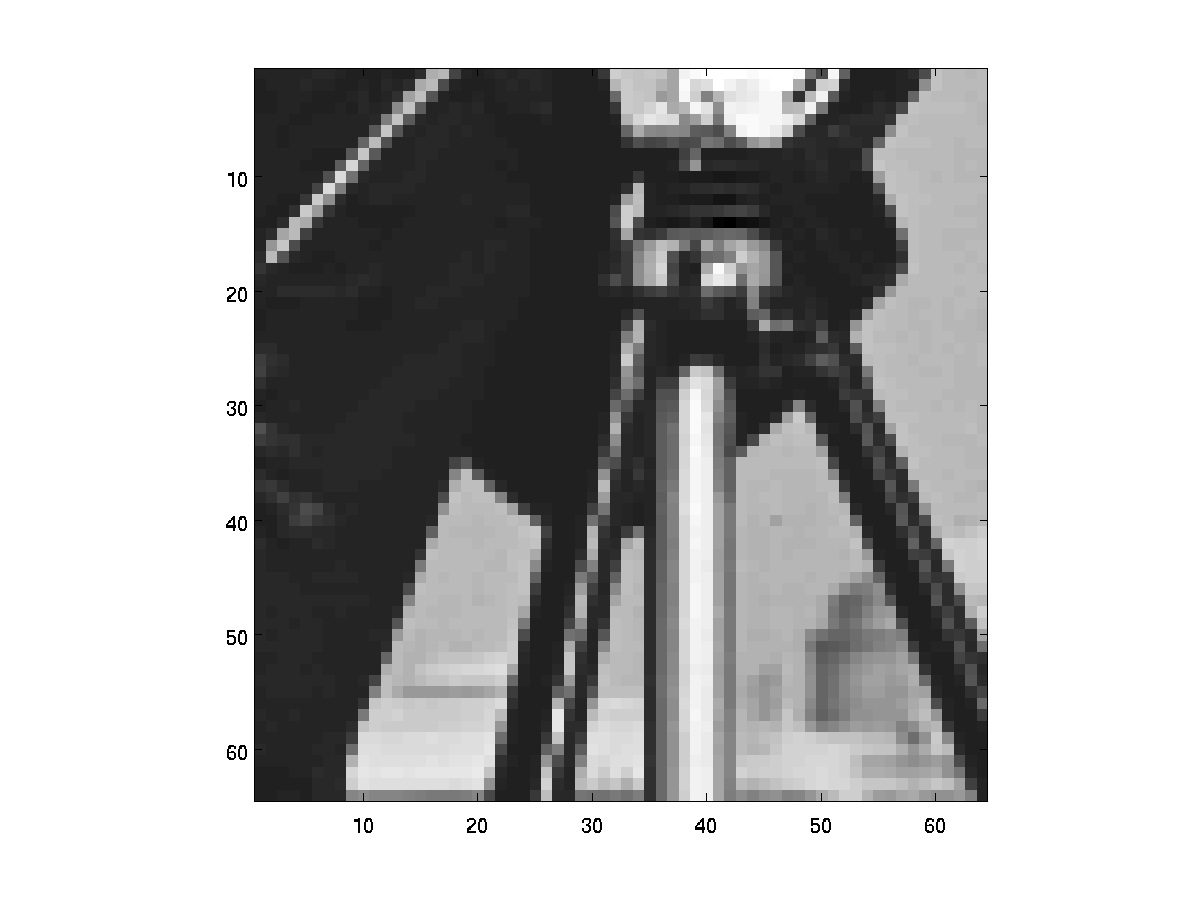}
    \label{fig:subcamancg}
}
\caption{The artifacts produced by a local minimizer of $\rho$.}
\label{fig:xlocmin}
\end{figure}

In order to find a local minimizer, we construct many random starting
guesses using the MATLAB {\tt rand} function.  To save time, we
chose to reconstruct a $64\times 64$ subimage of the cameraman image shown
in Figure~\ref{fig:caman}.  This subimage is shown in 
Figure~\ref{fig:subcaman}.  A $16 \times 16$ binary probe that has a
value 1 in the $8 \times 8$ center of the probe and 0 elsewhere is used.
The diffraction stack consisting of $64$ diffraction images is obtained 
by translating the probe 4 pixels a time in either the horizontal and 
vertical direction.

\begin{figure}[hbtp]
\centering
\hfill
\subfigure[Change of the relative residual norm (res).]{
    \includegraphics[width=0.45\textwidth]{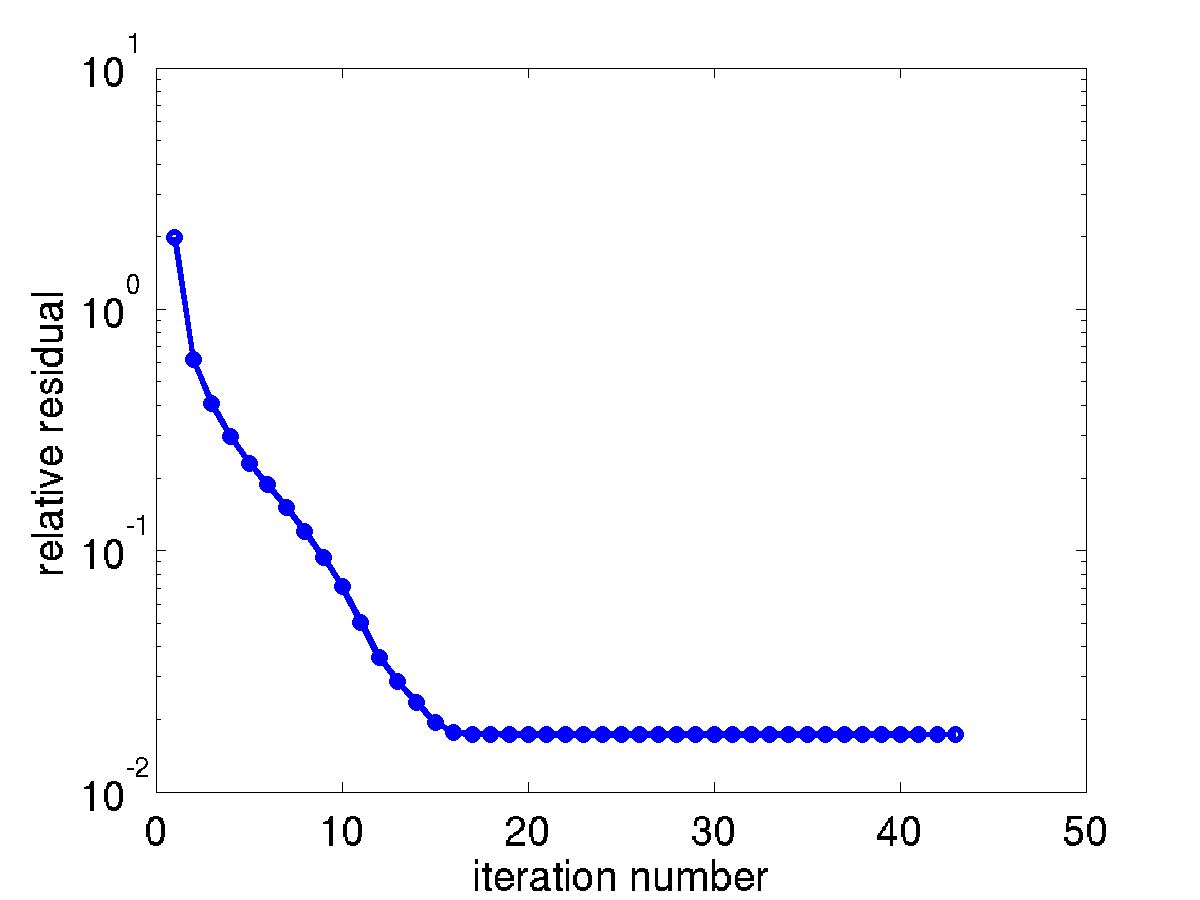}
    \label{fig:camanobjcg}
}
\hfill
\subfigure[Change of the relative gradient.]{
    \includegraphics[width=0.45\textwidth]{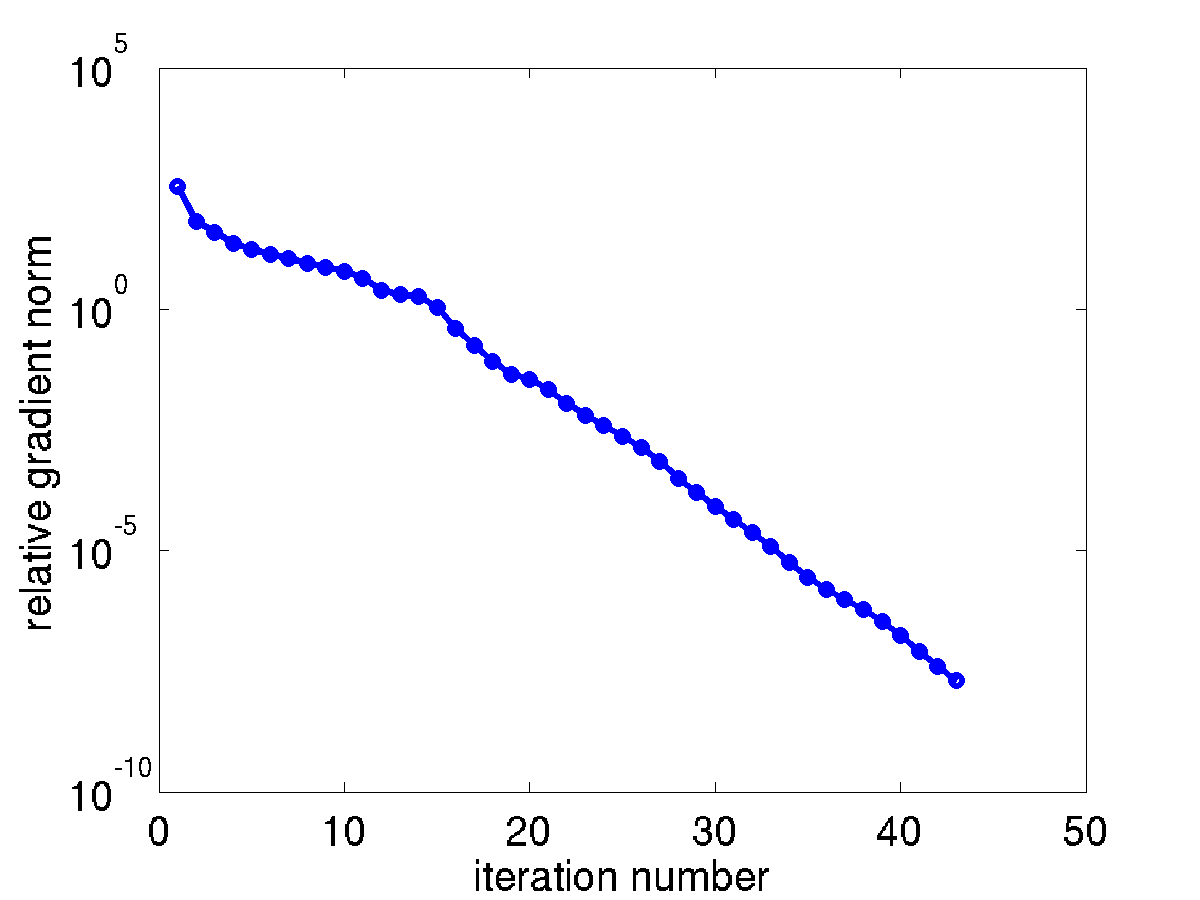}
    \label{fig:camangradcg}
}
\caption{The convergence of CG to a local minimizer.}
\hfill
\label{fig:cglocmin}
\end{figure}

Figure~\ref{fig:cglocmin} shows that one of the random starting 
guesses lead to the convergence of the CG algorithm to a local 
minimizer.  In particular, the relative residual (\ref{relres}) 
which is proportional to the objective function $\rho$ stagnates
around 0.9 after the first 15 iterations (Figure~\ref{fig:camanobjcg}), 
whereas the relative gradient $\| \nabla \rho(\psi^{(\ell)})\|/\|\hpsi\|$ 
decreases to $10^{-8}$ after 40 iterations. 

\begin{figure}[htbp]
  \centering
  \hfill
  \subfigure[Amplitude error in the reconstruct image]{
     \includegraphics[width=0.45\textwidth]{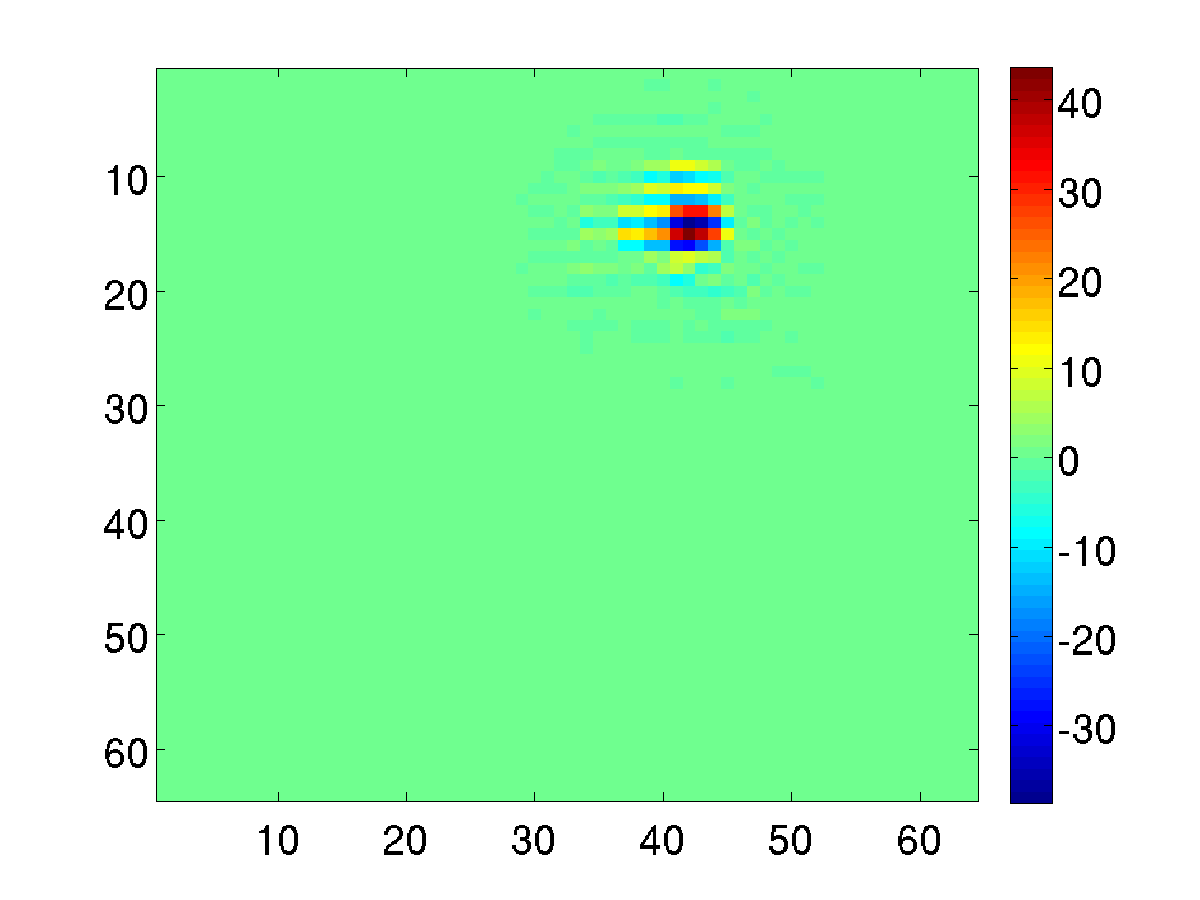}
     \label{fig:xlocerr}
  }
  \hfill
  \subfigure[Phase error in degrees associated with a particular frame]{
     \includegraphics[width=0.45\textwidth]{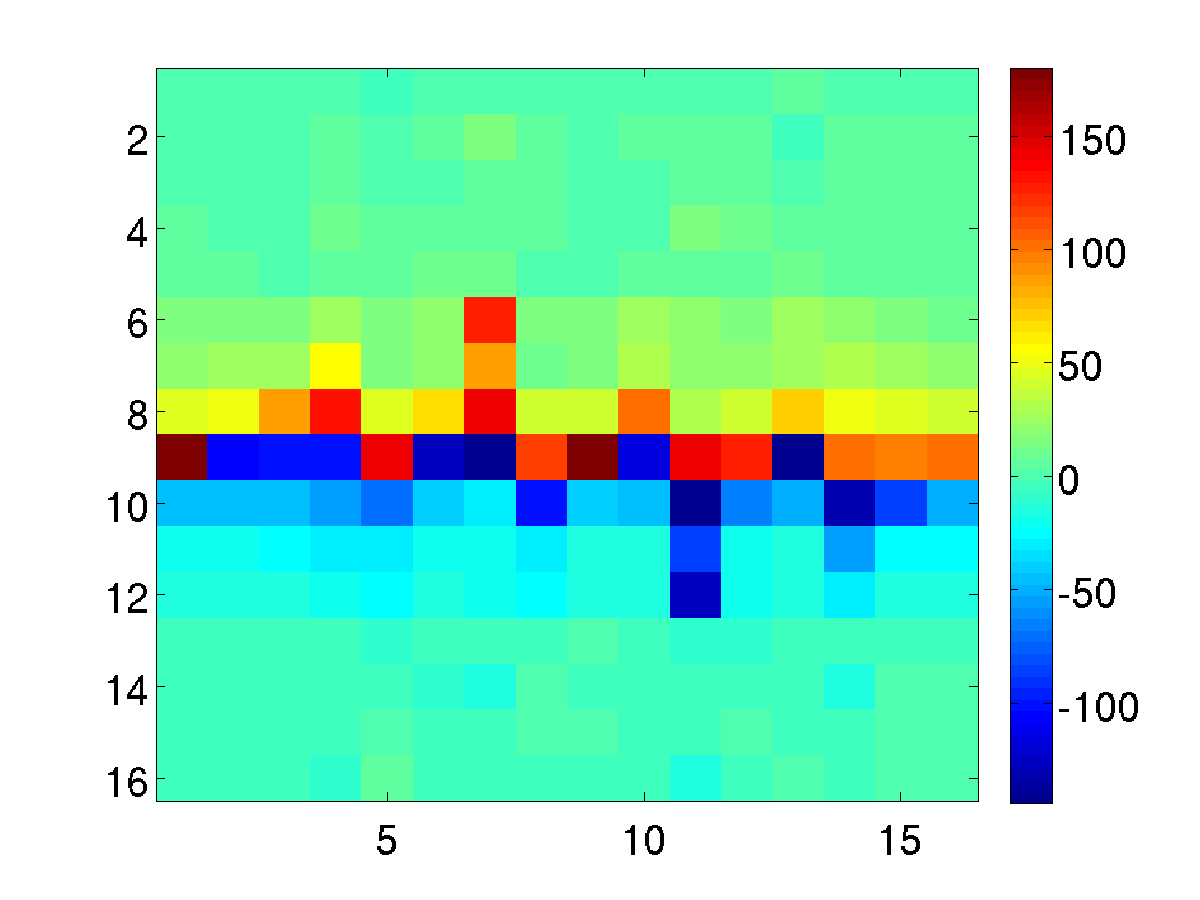}
     \label{fig:errface}
  }
\caption{The error associated with a local minimizer.}
\end{figure}

Figure~\ref{fig:subcamancg} shows how the reconstructed image compares
with the true image for this particular starting guess used.  In this
case, the local minimizer appears to contain visible artifacts in a small
region near top of the tripod. The amplitude of this localized error is 
also revealed in the relative error plot shown in Figure~\ref{fig:xlocerr}.
The phase error associated with a particular frame of the reconstruction
obtained from
\[
\frac{\overline{Q_i \psi}}{|Q_i \psi|} \cdot \frac{Q_i \hat{\psi}}{|Q_i \hat{\psi}|},
\] 
for some particular $Q_i$ is shown in Figure~\ref{fig:errface}.


We should also note that for this particular starting guess,
all methods we tried converged to the same local minmizer. This
is not all that surprising.  It simply shows (empirically) that 
local minimizers of (\ref{obj1}) exists, and our starting guess
is sufficiently close to it.

\begin{figure}[hbtp]
\centering
\hfill
\subfigure[The convergence of the NT algorithm when it is applied to (\ref{obj2})
           (red) and (\ref{wobj2}). The starting guess chosen in these runs
           is the same one used in the minimization of (\ref{obj1}).]{
    \includegraphics[width=0.45\textwidth]{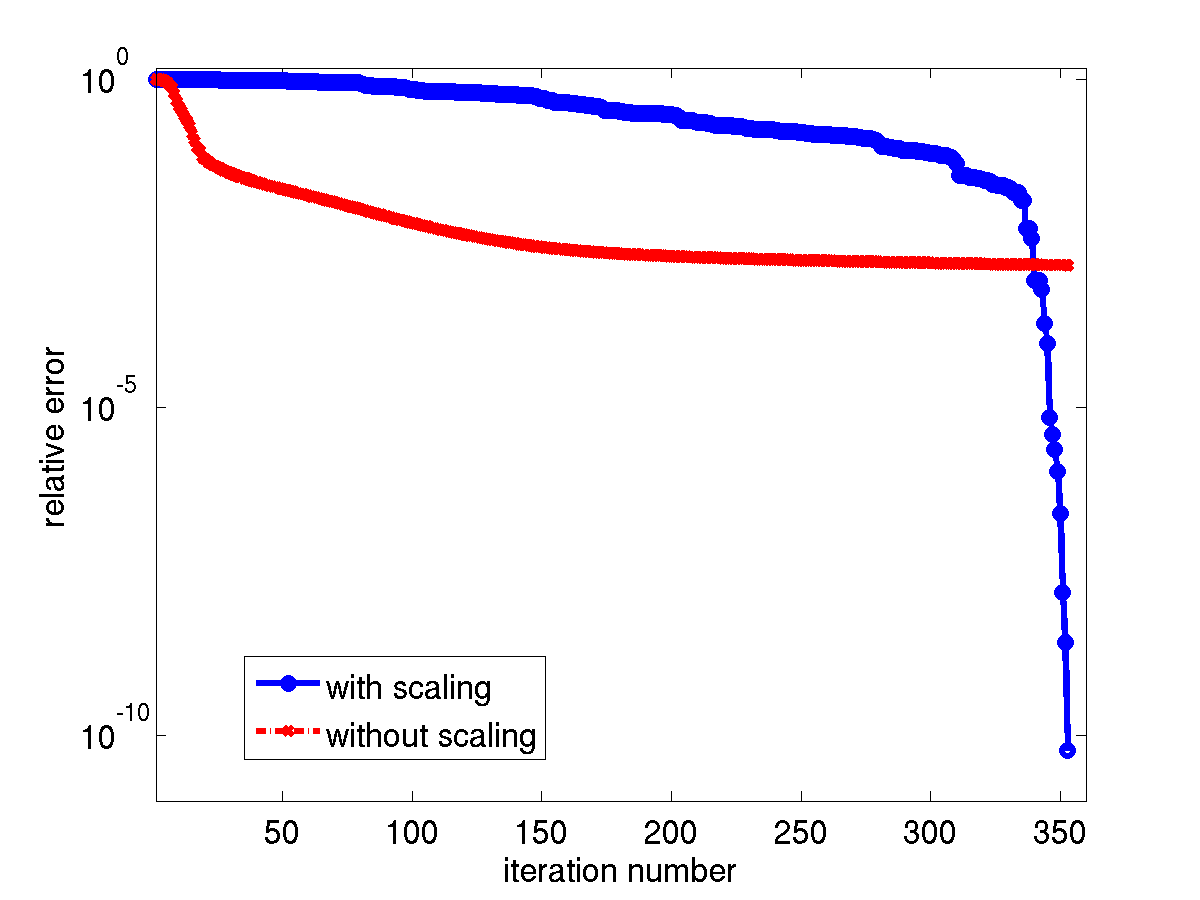}
    \label{fig:camepsconv1}
}
\hfill
\subfigure[The convergence of the NT algorithm when the starting guess
           is chosen to be the local minimizer shown in Figure~\ref{fig:subcamancg}]{
    \includegraphics[width=0.45\textwidth]{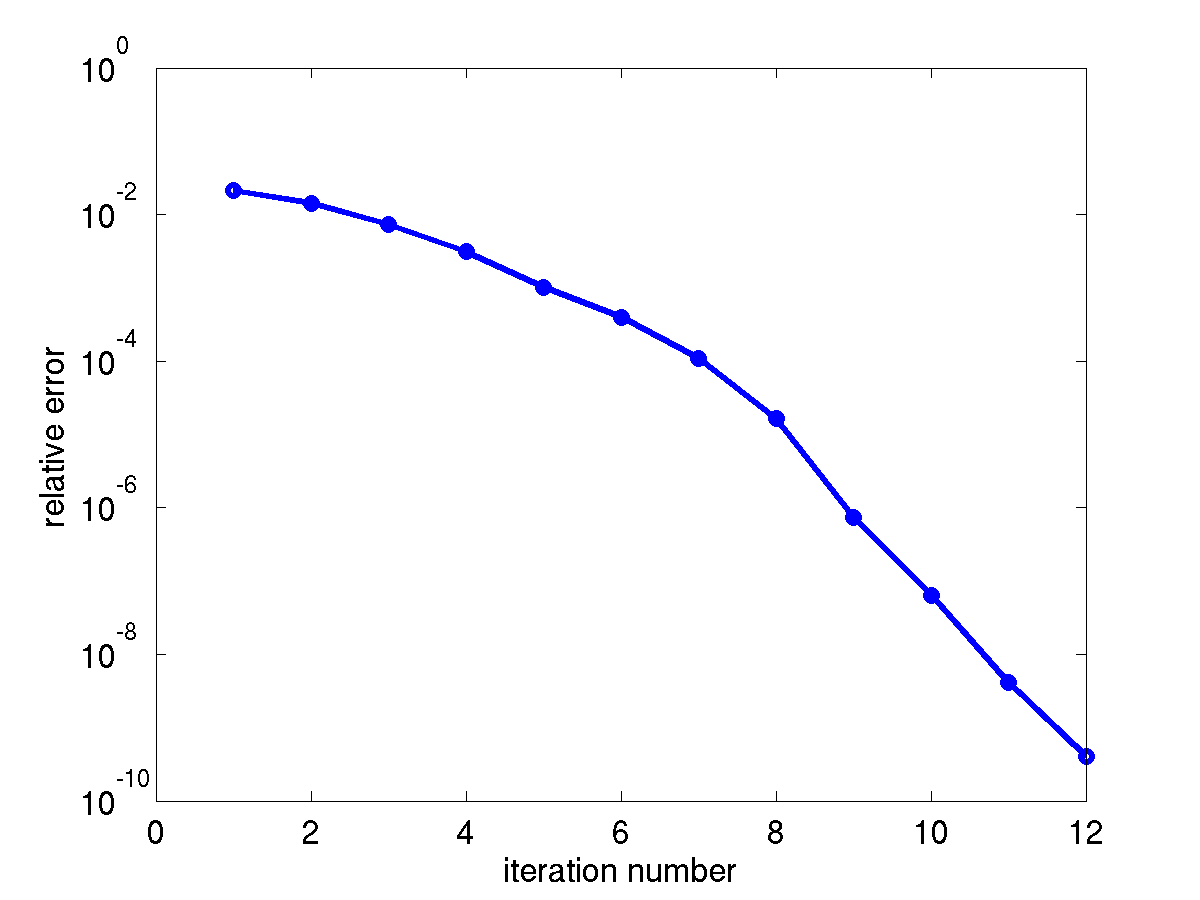}
    \label{fig:camepsconv2}
}
\caption{The convergence of the NT algorithm when applied to (\ref{obj2}) (red)
and (\ref{wobj2}) (blue).}
\label{fig:camepsconv}
\end{figure}

However, what is interesting is that if we choose to minimize
~(\ref{obj2}) by using any one of the iterative methods discussed
above from the same starting guess, we are able to obtain the
correct solution.  For examples, Figure~\ref{fig:camepsconv1} shows
that when the NT applied to the weighted (scaled) objective function
\begin{equation}
\tilde{\epsilon}_(\psi) = 
\frac{1}{2}\sum_{i=1}^{k} (|z_i|^2 - b_i^2)^T \Diag{b_i}^{-1}(|z_i|^2-b_i^2),
\label{wobj2}
\end{equation}
where $|z_i| = | FQ_i\psi |$ and $b_i = |FQ_i\hpsi|$, an accurate
reconstruction can be obtained in roughly 350 iterations.  
Admittedly, the convergence rate is much slower in this case when
compared to the convergence of NT when it's applied to (\ref{obj1}) 
from a different starting point.  The convergence is even slower 
if no weighting (or scaling) is used, i.e. when (\ref{obj2}) is used
as the objective function.  However, the fact that convergence can be 
reached for (\ref{wobj2}) but not (\ref{obj1}) from the same starting point 
is quite interesting.  Furthermore, Figure~\ref{fig:camepsconv2} 
shows that if we take the local minimizer returned from an iterative
minimization of (\ref{obj1}) as the starting guess for minimizing
(\ref{wobj2}), convergence can be reached in 12 iterations.  
This experiment suggests that it may be useful to have a hybrid 
optimization scheme in which (\ref{obj1}) is minimized first. If 
a local minimizer of (\ref{obj1}) is identified, one can then try
to minimize (\ref{wobj2}) starting from the local minimizer of 
(\ref{obj1}).

\subsection{The Effect of Overlapping on the Convergence of Iterative
Algorithm} \label{sec:trans}
As we alluded to earlier, the amount of overlap between two adjacent 
diffraction frames has a noticeable effect on the convergence of 
optimization based iteration algorithms (e.g., CG, NT, SD etc.) used 
to reconstruct the true image.  Although we currently do not have a 
clear way to quantify such an effect mathematically, the following 
examples demonstrate this effect.

\begin{figure}[htbp]
\hfill
\subfigure[The effect of overlapping on the convergence of CG for
           the gold ball image reconstruction.]{
    \includegraphics[width=0.45\textwidth]{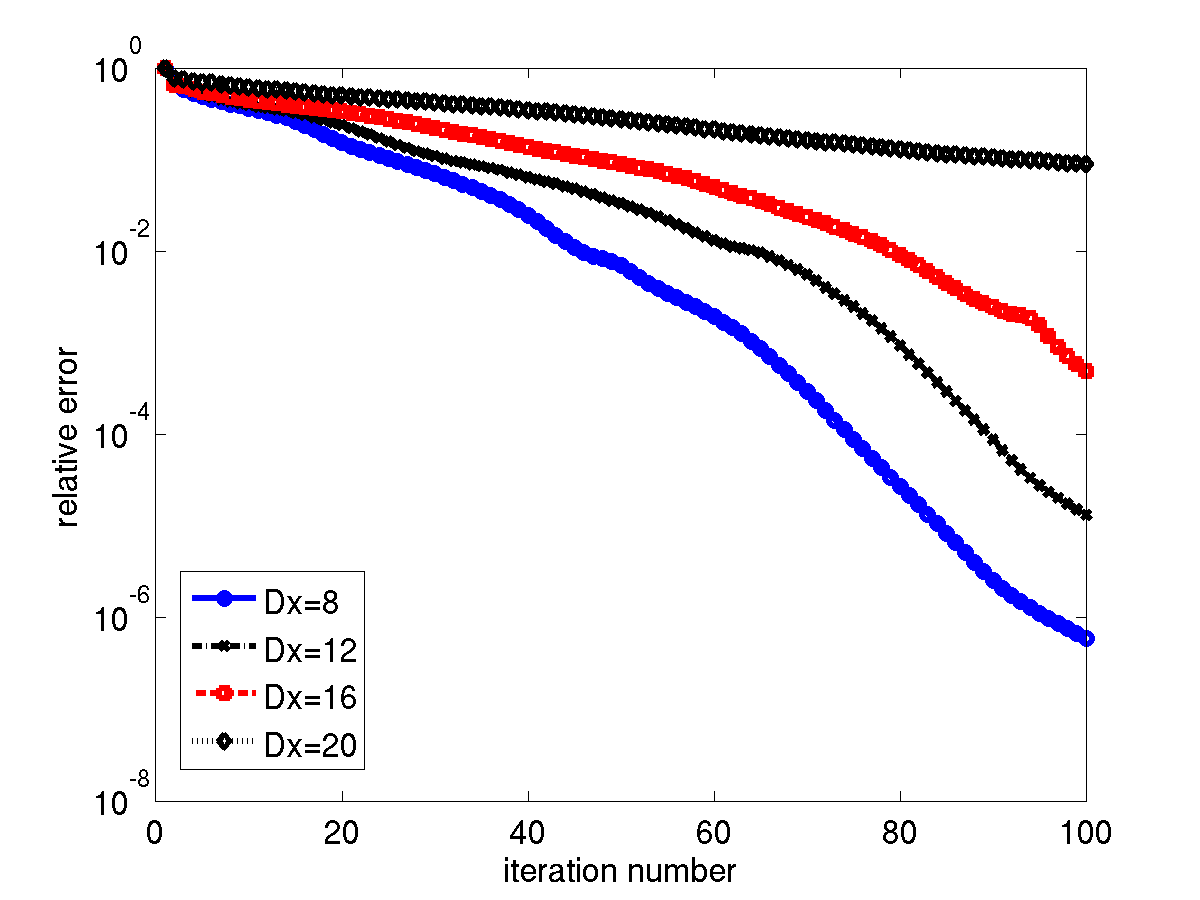}
    \label{fig:audxcg}
}
\hfill
\subfigure[The effect of overlapping on the convergence of HIO for
           the gold ball image reconstruction.]{
    \includegraphics[width=0.45\textwidth]{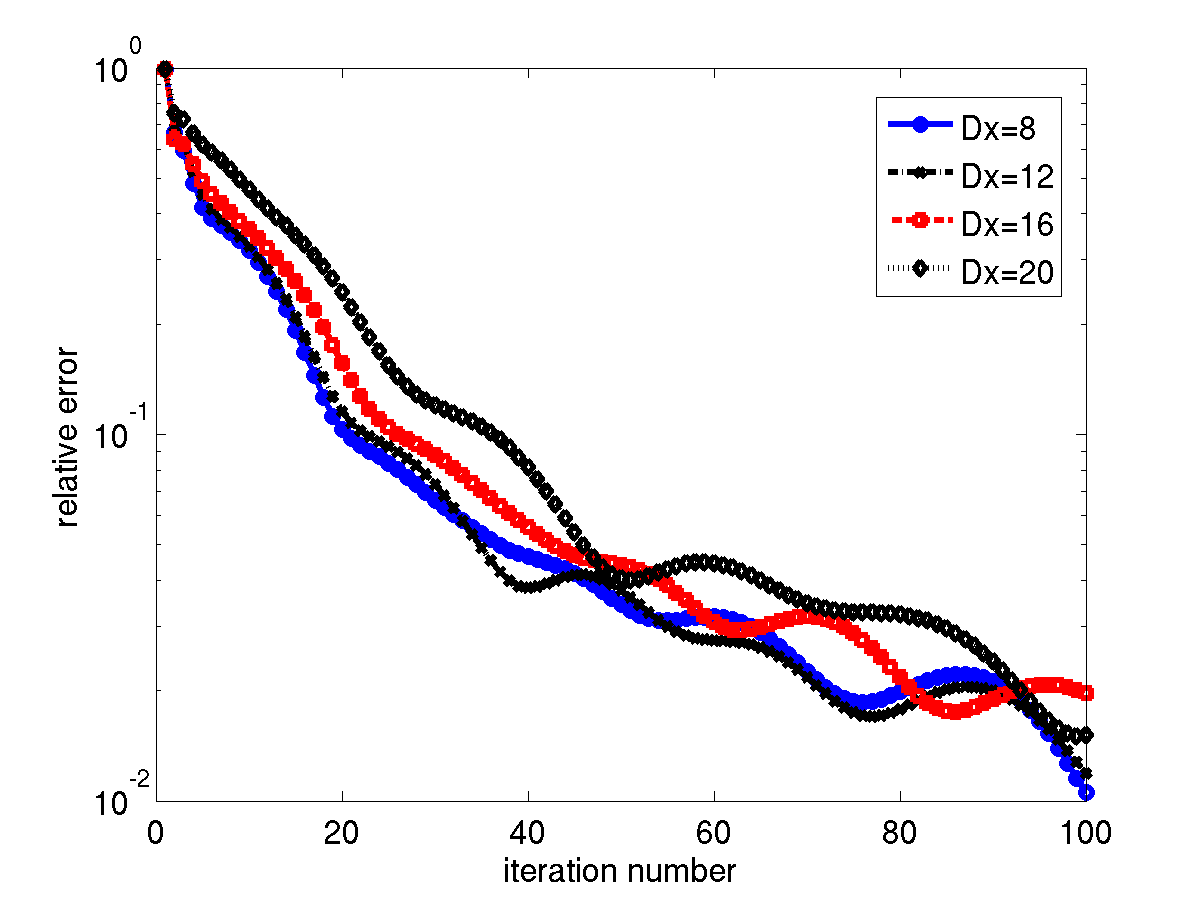}
    \label{fig:audxhio}
}
\caption{The effect of overlapping on the convergence of CG and HIO algorithms.}
\label{fig:audx}
\end{figure}

In the first example, we try to reconstruct the gold ball image from 
four different diffraction stacks. Each stack contains a set of 
$64\times 64$ diffraction frames.  These frames are generated by
translating the probe shown in Figure~\ref{fig:maskau} by different
amount in horizontal and vertical directions.  The larger the translation,
the smaller the overlap is between two adjacent images.
Figure~\ref{fig:audxcg} shows that CG converges very slowly when 
the diffraction stack contains diffraction frames obtained by
translating the probe $20$ pixels at a time (the black curve). 
Faster convergence is observed when the amount of 
translation is decreased to $\Delta x = 16,12,8$.  It is interesting
to see from Figure~\ref{fig:audxhio} that the amount of overlap
does not affect the convergence of the HIO algorithm.

In the second example, we try to reconstruct the gold ball image from
1024 diffraction frames of $128\times 128$ pixels.  
The illumination function is similar to that used in Figure~\ref{fig:aumask}.
It is scaled by a factor of 2 to $128\times 128$ pixels. The probe FWHM
(full width at half maximum) is 30 pixels.  We choose to fix the number of 
frames. So the reconstructed area increases with step size.  
When probe is near the edge of the image, we ``wrap it around the edge"
as if the image itself is periodically extended.
The overlap is varied by changing the step size $\Delta x$. The larger the
$\Delta x$, the smaller the amount of overlap.

\begin{figure}[hbtp]
  \begin{center}
    \subfigure[convergence of CG from 100 random starts $\Delta x=20$]{
    \includegraphics[width=0.45\textwidth]{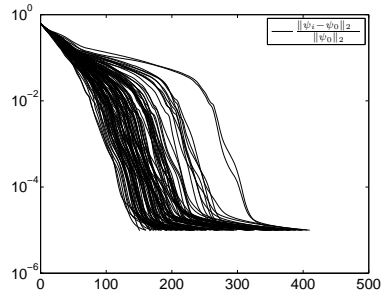}
    \label{fig:cgtests}
    }
\hfill
\subfigure[percentage tests that converge to $err \le10^{-4}$
]{
    \includegraphics[width=0.45\textwidth]{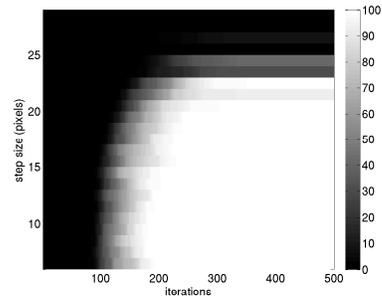}
    \label{fig:cgfails}
}
    \subfigure[convergence of RAAR from 100 random starts $\Delta x=20$]{
    \includegraphics[width=0.45\textwidth]{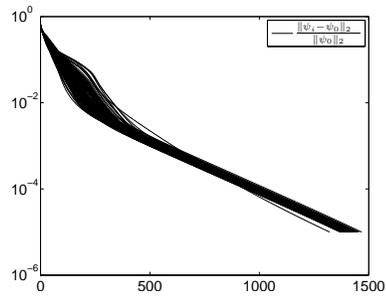}
    \label{fig:raartests}   
    }
\hfill
\subfigure[percentage of RAAR iterations that converge to an error of $10^{-4}$
]{
    \includegraphics[width=0.45\textwidth]{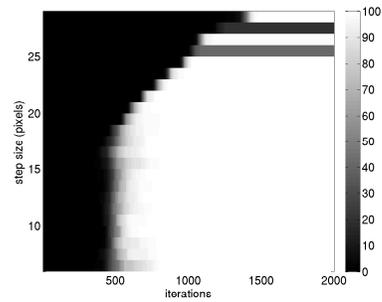}
    \label{fig:raarfails}
}

  \end{center}
  \caption{The convergence rate of the CG and RAAR methods from
    different random starting points. 
    }
  \label{fig:convergence}
\end{figure}

The starting point is produced from a random number generator for each
test.  A range of step sizes between 6 and 30 pixels have been tried. 
For a fixed step size, the test is repeated 100 times. 
We observe that the step size $\Delta x$ does not influence the 
convergence rate up to $\Delta x\simeq 20$.  Figures~\ref{fig:cgtests} and
~\ref{fig:raartests} show that the conjugate gradient method converges in 
less than 400 iterations, while the RAAR algorithm requires almost 1500 
iterations.  Figures~\ref{fig:cgfails} and~\ref{fig:raarfails} illustrate
the percentage of successful runs started from a random guess 
for each of the step sizes $0\leq \Delta x \leq 30$. The percentage of 
successful runs (shown in color) is plotted against the maximum number of 
allowed iterations.  When $\Delta x \leq 20$, both CG and RAAR 
converge nearly 100\% of the time when a relatively small number of 
iterations are used in these methods.  However, when 
$20 \leq \Delta x \leq 25$, more iterations are required to ensure
the convergence of CG and RAAR. When $25 \leq \Delta x \leq 30$, CG
appears to stagnate for all random starting guesses we tried, whereas
RAAR can still converge when a very large number of iterations are taken.

To explain the effect of overlapping on the convergence of 
optimization based iterative algorithms such as the nonlinear CG,
we examine the structure of the Hessian of the objective function
$\rho$ in (\ref{obj1}).  It follows from (\ref{hpsipsi2})-(\ref{hbpsibpsi2}) 
that the $H^\rho$ can be written as
\begin{equation}
H^{\rho} =  
\left( (\hF Q)^{\ast} \ \ (\hF Q)^T \right)
\left(
\begin{array}{cc}
B_{11} & B_{12} \\
B_{21} & B_{22}
\end{array}
\right)
\left(
\begin{array}{c}
\hF Q \\
\bar{\hF}\bar{Q}
\end{array}
\right),
\label{bhess}
\end{equation}
where $B_{11} = \overline{{B}_{22}}$ and $B_{12} = B_{21}^{\ast}$ are all
diagonal, $\hat{F}$ is a block diagonal matrix of discrete Fourier transforms, 
i.e. $\hat{F} = \Diag{F,F,...,F}$, and 
$Q = (Q_1^{\ast} \: Q_2^{\ast} \: ... \: Q_k^{\ast})^{\ast}$.
The diagonal elements of $B_{11}$ and $B_{12}$ are simply 
$1 - \beta_{ji}/(2\zeta_{ji})$ and 
$\beta_{ji}\zeta_{ji}^2/(2 |\zeta_{ji}|^3)$ respectively 
for $i = 1,2,...,k$ and $j = 1, 2, ..., m$. 

We will show that $H^\rho$ is diagonal dominant when there is a sufficient
amount of overlap between adjacent diffraction frames.
To simplify our discussion, let us assume for the moment that
$b_i$ is a 1D diffraction pattern obtained from a binary probe that 
illuminates three pixels at a time, and the probe is translated one pixel 
at a time so that the image frame that produces $b_i$ overlaps with 
that produces $b_{i-1}$ by two pixels. In this case, the $\hat{F}Q$ term in 
(\ref{bhess}) has the form
\[
\left(
\begin{array}{ccccc}
f_1 & f_2 & f_3 & \hdots  &  0 \\
 0  & f_2 & f_3 & \ddots  & \vdots \\
 0  & 0   & f_3 & \ddots  & f_k \\   
f_1 & 0   & 0   & \ddots  & f_k \\
f_1 & f_2 & 0   & \hdots  & f_k 
\end{array}
\right),
\]
where $f_i$ is the $i$th column of $F$.

As a result, a typical diagonal term of $H^\rho$ has the
form
\begin{equation}
H^{\rho}_{i,i} = f_i^{\ast}D_{i-2}f_i + f_i^{\ast}D_{i-1}f_i 
+ f_i^{\ast} D_i f_i = \mbox{trace}(D_{i-2} + D_{i-1} + D_i),
\label{Hdiag}
\end{equation}
where $D_i$ is a diagonal matrix that contains elements $1 - \beta_{ji}/(2\zeta_{ji})$ for $j = 1, 2, 3$.

When $\psi$ is near the solution, $z_i$ is close to $b_i$. Hence,
$D_i$ is likely to contain positive entries only.  Therefore,
the diagonal elements of $H^{\rho}$ are likely to be much larger
compared to the nonzero off-diagonal elements which contain
terms in the form of either $f_j^{\ast} D_i f_{\ell}$ and its
conjugate, where $j \neq \ell$, or  $f_j^T E_i f_{\ell}$
and its conjugate, where $E_i$ is a diagonal matrix (and part of $B_{12}$) that
contains elements $\beta_{ji}\zeta_{ji}^2/(2 |\zeta_{ji}|^3)$ for
$j = 1,2,3$.  Due to the phase difference between 
$f_j$ and $f_{\ell}$, $D_i$'s do not add up ``coherently" on 
the off-diagonal of $H^{\rho}$ as they do on the diagonal. 
Neither do nonzero entries in $E_i$'s add up coherently on the off-diagonal 
blocks of $H^\rho$ either.  Hence, the matrix $H^{\rho}$ becomes 
diagonal dominant when there is larger amount of overlap
between two adjacent frames. In fact, the diagonal of $H^{\rho}$ 
may become so dominant that the spectral property of $H^{\rho}$ is determined 
largely by the diagonal part of the matrix, which is typically well 
conditioned due to the averaging of $D_i$ in (\ref{Hdiag}). This 
observation provides an intuitive explaination on why increasing the 
amount of overlap between adjacent frames tends to improve 
the convergence rate of CG and other optimization based iterative 
ptychographical phase retrieval algorithms.  Although this is not a 
precise analysis of the spectral property of $H^{\rho}$, the analysis
does match with observations made in our numerical expriments.  Moreover, 
this type of analysis can be extended to the 2D case in which $F$ is 
represented as a tensor product of two 1D discrete Fourier transforms.
 
%

\section{Conclusion}
We formulated the ptychographic phase retrieval problem as 
a nonlinear optimization problem and discussed how standard
iterative optimization algorithms can be applied to solve this 
problem.  

We showed that the optimization problems we solve
are not globally convex. Hence standard optimization algorithms
can produce local minimizers.  However, the Hessian of the 
objective functions we minimize do have special structures that 
may be exploited.  

We compared the performance of several 
optimization algorithms and found that Newton's method with
Steihaug's trust region technique gave the best performance
on a real valued image.  For a complex valued image, the 
nonlinear conjugate gradient algorithm appears to perform better. 

We discussed the effect of preconditioning on convergence of the 
CG algorithm.  We also demonstrated it is possible for an optimization 
algorithm to converge to a local minimizer although in practice
such type of convergence failure is rare, especially when 
the amount of overlap between two adjacent diffraction frames
is large.

We demonstrated by a numerical example that the convergence
rate of an optimization algorithm depends on the amount of
overlapping between two adjacent diffraction frames. We 
provided an intuitive analysis on why this occurs.
More research is needed to provide a more precise analysis
on this phenonmenon.

We identified the connection between the optimization based 
approach with both Wigner deconvolution and projection algorithms
often used in phase retrieval literatures. We pointed out the 
limitation of Wigner deconvolution and showed that the optimization
based algorithm tend to perform better than projection algorithms
such as HIO when the amount of overlap between adjacent images
is sufficiently large.

\section*{Acknowledgment}
This work was supported by the Laboratory Directed Research and
Development Program of Lawrence Berkeley National Laboratory under
the U.S. Department of Energy contract number DE-AC02-05CH11231 (C. Y.,
A. S., S. M.), the National Science Foundation Grant 0810104 (J. Q.)
and by the Director, Office of Science, Advanced Scientific Computing Research,
of the U.S. Department of Energy under Contract No. DE-AC02-05CH11231 (F.M.).
The computational results presented were obtained at the National Energy
Research Scientific Computing Center (NERSC), which is supported by
the Director, Office of Advanced Scientific Computing Research of the
U.S. Department of Energy under contract number DE-AC02-05CH11232.

\end{document}